\documentclass[a4paper,12pt]{article}
\pdfoutput=1
\usepackage{graphicx, rotating,amssymb,amsmath}
\usepackage{cite}

\ifx\pdfoutput\undefined
\usepackage[dvips,bookmarks]{hyperref}	% This is for arXiv.org
\else
\usepackage{hyperref}	% This is for pdftex
\fi
\hypersetup{colorlinks,bookmarksopen,bookmarksnumbered,citecolor=verdeCP3,
linkcolor=bluCP3,pdfstartview=FitH,urlcolor=rossoCP3}

\def\hhref#1{\href{http://arxiv.org/abs/#1}{#1}} % in bibliography
\usepackage{multicol}
\usepackage{color}
\definecolor{rosso}{cmyk}{0,1,1,0.4}
\definecolor{rossos}{cmyk}{0,1,1,0.55}
\definecolor{rossoc}{cmyk}{0,1,1,0.2}
\definecolor{blu}{cmyk}{1,1,0,0.3}
\definecolor{blus}{cmyk}{1,1,0,0.6}
\definecolor{bluc}{cmyk}{1,1,0,0.1}
\definecolor{verde}{cmyk}{0.92,0,0.59,0.25}
\definecolor{verdec}{cmyk}{0.92,0,0.59,0.15}
\definecolor{verdes}{cmyk}{0.92,0,0.59,0.4}

\font\tenrsfs=rsfs10 at 12pt
\font\sevenrsfs=rsfs7
\font\fiversfs=rsfs5
\newfam\rsfsfam
\textfont\rsfsfam=\tenrsfs
\scriptfont\rsfsfam=\sevenrsfs
\scriptscriptfont\rsfsfam=\fiversfs
\def\mathscr#1{{\fam\rsfsfam\relax#1}}

\def\baselinestretch{0.97}

\def\circa#1{\,\raise.3ex\hbox{$#1$\kern-.75em\lower1ex\hbox{$\sim$}}\,}

\newcommand{\beq}{\begin{equation}}
\newcommand{\eeq}{\end{equation}}

\def\circa#1{\,\raise.3ex\hbox{$#1$\kern-.75em\lower1ex\hbox{$\sim$}}\,}
\makeatletter

\def\art{\@ifnextchar[{\eart}{\oart}}
\def\eart[#1]#2#3#4#5#6{{\rm #2}, {#3 #4} {\rm (#6) #5} [{\hhref{#1}}]}
\def\hepart[#1]#2{{\rm #2, \hhref{#1}}}
\newcommand{\oart}[5]{{\rm #1}, {#2 #3} {\rm (#5) #4}}

\newcounter{alphaequation}[equation]
\def\thealphaequation{\theequation\hbox to
0.6em{\hfil\alph{alphaequation}\hfil}}
\def\eqnsystem#1{
\def\@eqnnum{{\rm (\thealphaequation)}}
\def\@@eqncr{\let\@tempa\relax \ifcase\@eqcnt \def\@tempa{& & &} \or
  \def\@tempa{& &}\or \def\@tempa{&}\fi\@tempa
  \if@eqnsw\@eqnnum\refstepcounter{alphaequation}\fi
\global\@eqnswtrue\global\@eqcnt=0\cr}
\refstepcounter{equation} \let\@currentlabel\theequation \def\@tempb{#1}
\ifx\@tempb\empty\else\label{#1}\fi
\refstepcounter{alphaequation}
\let\@currentlabel\thealphaequation
\global\@eqnswtrue\global\@eqcnt=0 \tabskip\@centering\let\\=\@eqncr
$$\halign to \displaywidth\bgroup \@eqnsel\hskip\@centering
$\displaystyle\tabskip\z@{##}$&\global\@eqcnt\@ne
\hskip2\arraycolsep\hfil${##}$\hfil& \global\@eqcnt\tw@\hskip2\arraycolsep
$\displaystyle\tabskip\z@{##}$\hfil
\tabskip\@centering&\llap{##}\tabskip\z@\cr}
\def\endeqnsystem{\@@eqncr\egroup$$\global\@ignoretrue} \makeatother

\long\def\symbolfootnote[#1]#2{\begingroup\def\thefootnote{\fnsymbol{footnote}}
\footnote[#1]{#2}\endgroup}

\definecolor{rossoCP3}{cmyk}{0,.88,.77,.40}
\definecolor{verdeCP3}{rgb}{0.09765625, 0.57421875, 0.1015625}
\definecolor{bluCP3}{rgb}{0, 0.23, 0.67}

\begin{document}

\begin{titlepage}
 \begin{center}
{{\LARGE {\color{rossoCP3}
\bf  Light Magnetic Dark Matter  in \\ Direct Detection Searches \rule{0pt}{25pt}}
}} 
 \end{center}
 \par \vskip .2in \noindent
\begin{center}
{\sc {\color{black}Eugenio Del Nobile\!\!\symbolfootnote[1]{delnobile@cp3-origins.net}, Chris Kouvaris\!\!\symbolfootnote[2]{kouvaris@cp3-origins.net}, Paolo Panci\!\!\symbolfootnote[3]{panci@cp3-origins.net}, \\ Francesco Sannino\!\!\symbolfootnote[4]{sannino@cp3.dias.sdu.dk}, Jussi Virkaj\"arvi\!\!\symbolfootnote[7]{virkajarvi@cp3-origins.net}}
}
\end{center}
\begin{center}
  \par \vskip .1in \noindent
\mbox{\it
CP$\,^3$-Origins and DIAS,
University of Southern Denmark,
Odense, Denmark}
   \par \vskip .5in \noindent
\end{center}
\begin{center}{\large Abstract}\end{center}
\begin{quote}
We study a fermionic Dark Matter particle carrying magnetic dipole moment and analyze its impact on direct detection experiments. In particular we show that it can accommodate the DAMA, CoGeNT and CRESST experimental results. Assuming conservative bounds, this candidate is shown not to be ruled out by the CDMS, XENON and PICASSO experiments. We offer an analytic understanding of how the long-range interaction modifies the experimental allowed regions, in the cross section versus Dark Matter mass parameter space, with respect to the typically assumed contact interaction. Finally, in the context of a symmetric Dark Matter sector, we determine the associated thermal relic density, and further provide relevant constraints imposed by indirect searches and colliders. 
\\
[.33cm]
{%\footnotesize
\small \it {Preprint: CP$\,^3$-Origins-2012-007 \& DIAS-2012-8}}
 \end{quote}
\par \vskip .1in
\vfill
 \end{titlepage}

         \newpage

\def\baselinestretch{1.0}
\tiny
\setlength{\unitlength}{1mm}
%\begin{fmffile}{diagrammi}

%\noindent{\bf \vvv{green} = new}
%\\
%{\bf \rrr{red} = corrections}
%\\
%{\bf \colorbox{yellow}{highlighted} = to be implemented}

\hspace{20mm}
\normalsize

\tableofcontents

\newpage

\section{Introduction}

The presence of a relatively big dark component in the matter content of the Universe has been now established, and part of the scientific community is trying to unravel its still enigmatic properties. Despite the conspicuous number of experiments that probed the gravitational effects of this Dark Matter (DM) on the known matter, we still lack a more direct evidence of what it actually is and what are its properties.

Under the assumption that the DM is constituted by one or more unknown particles, several experiments have attempted recently to probe its microscopic nature and its interactions. Different search strategies are possible. For indirect detection experiments, traces of DM annihilations to known particles are looked for in cosmic rays or in the spectrum of solar neutrinos. Collider searches are also performed, in case the DM can be produced in sizable amounts to have an impact in the channels featuring missing energy. Direct detection experiments rely instead on direct scatterings of DM particles off nuclei of ordinary matter, inside shielded detectors placed underground in order to diminish the cosmic rays background.

DM direct detection experiments are providing exciting results. For example the \mbox{DAMA/NaI} collaboration \cite{Bernabei:1998fta} has claimed to have observed the expected annual modulation of the DM induced nuclear recoil rate \cite{Drukier:1986tm, Freese:1987wu}, due to the rotation of the Earth around the Sun. The rotation, in fact, causes a different value of the flux of DM particles hitting the Earth depending on the different periods of the year. The upgraded DAMA/LIBRA detector has further confirmed \cite{Bernabei:2008yi} the earlier result adding much more statistics, and it has reached a significance of $8.9 \sigma$ C.L. for the cumulative exposure \cite{Bernabei:2010mq}. 

Interestingly the DAMA annual modulation effect has been shown to be compatible with a DM interpretation which, for the case of coherent spin-independent scattering, leads to a range of DM masses spanning from a few GeV up to a few hundred GeV, and cross sections between $10^{-42}$ cm$^2$ to $10^{-39}$ cm$^2$ \cite{Bernabei:1998fta,Bernabei:2008yi,Bernabei:2010mq}, with some noticeable differences due to the galactic halo modeling \cite{Belli:2002yt,Belli:2011kw}.

More recently, the CoGeNT experiment first reported an irreducible excess in the counting rate \cite{Aalseth:2010vx}, which could also be in principle  ascribed to a DM signal. In the last months, the same experiment reported an additional analysis which shows that the time-series of their rate is actually compatible with an annual modulation effect \cite{Aalseth:2011wp}. The evidence of such a modulation for CoGeNT is at the level of $2.8 \sigma$ C.L.

Also the CRESST collaboration observed an excess \cite{Angloher:2011uu}. In particular, 67 counts were found in the DM acceptance region, where the estimated background from leakage of $e$/$\gamma$ events, neutrons, $\alpha$ particles and recoiling nuclei in $\alpha$ decays, is not sufficient to account for all the observed events. The analysis made by the collaboration rejects the background-only hypothesis at more than $4 \sigma$ \cite{Angloher:2011uu}.

The interesting feature is that the DAMA and CoGeNT results appear to be compatible for relatively light DM particles, in the few GeV to tens of GeV mass range and coherent scattering cross section around $10^{-40}$ cm$^2$. CRESST points somehow to larger DM masses, but it is still compatible with the range determined by the other two experiments. The actual relevant range of masses and cross sections depends on assumptions of the galactic DM properties, namely the velocity distribution function and the local DM density \cite{Belli:2011kw}. Other relevant analyses can be found in Refs. \cite{Foot:2011pi,Schwetz:2011xm,Farina:2011pw,McCabe:2011sr,Fox:2011px,Hooper:2011hd,Gondolo:2011eq,DelNobile:2011je,Arina:2011si,Frandsen:2011ts,Kaplan:2011yj,Feng:2011vu,Fitzpatrick:2010br,Hooper:2010uy,Foot:2010rj,Chang:2010yk,Fitzpatrick:2010em,Kopp:2009qt}. 

The CDMS and XENON100 experiments have recently reported a small number of events which pass all the selection cuts. Specifically, they have 2 events for CDMS \cite{Ahmed:2009zw} and 6 events for XENON100, reduced to 3 events after post-selection analysis \cite{Aprile:2011hi}, which are still too few to be interpreted as potential DM signal. They can therefore provide upper bounds on the DM scattering cross section. These constraints seem to set severe bounds on the DM parameter space, and therefore to rule out much of the allowed regions of the other experiments. Very recently the PICASSO experiment \cite{Archambault:2012pm} seems to provide even more stringent constraints for the low mass DM interpretation of DAMA, CoGeNT and CRESST.

However, there are at least two caveats when interpreting the results from the experiments mentioned above.
The first is that one has to pay attention to the fine details associated to the results quoted by each experiment, since a number of factors can affect the outcome. For example the actual response of the XENON and CDMS detectors can be uncertain and model dependent for a low energy signal \cite{Bernabei:2008jm,Collar:2011wq,Collar:2011kf}. Moreover, for crystal detectors like DAMA, an additional source of uncertainty is provided by the presence of an unknown fraction of nuclear recoils undergoing channeling. This effect is currently being investigated \cite{Bernabei:2007hw, Bozorgnia:2010xy, Feldstein:2009np}.  If confirmed, it would lead  to a significant shift of the DAMA allowed regions in the DM parameter space. Another source of uncertainty is associated with the details of the nuclei form factors for each experiment.

The second caveat is associated to the interpretation of the actual data within a very simple-minded model of the DM interaction with nuclei. For example, for the spin-independent case it is often assumed that DM couples via a contact interaction with equal strength to the proton and neutron. Upon relaxing some of these assumptions, less stringent constraints can be drawn.

\bigskip

In this work we perform a thorough investigation of the effects of a magnetic dipole moment of the DM on the direct detection experiments. We refer to \cite{Bagnasco:1993st,Pospelov:2000bq, Fitzpatrick:2010br, Sigurdson:2004zp, Barger:2010gv, Chang:2010en, Cho:2010br, Heo:2009vt, Gardner:2008yn, Masso:2009mu, Banks:2010eh, Fortin:2011hv, Barger:2012pf, Cline:2012is} for a limited sample of the earlier literature. 

A fermionic DM particle with a magnetic dipole moment can arise for example in models of composite DM \cite{Foadi:2007ue,Ryttov:2008xe,Sannino:2008nv}, or more generally in models featuring a new strong interaction (see e.g.~\cite{Sannino:2009za}). The application to cosmic rays as well as a complete discussion also of the light asymmetric DM paradigm, and last, but not the least, the first link to the study of grand unifications using DM appeared in \cite{Nardi:2008ix}. Recently it was shown, for the first time, via first principle lattice simulations \cite{Lewis:2011zb} that one has, indeed, for minimal models of Technicolor the correct pattern of chiral symmetry breaking leading to natural candidates of light composite DM. Despite the possibility to link the DM issue to other unresolved fundamental questions like e.g.~the (dynamical) breaking of the Electroweak symmetry or to the grand unification problem,  we investigate this DM candidate in a model independent way addressing only its DM phenomenology. The main feature of the present DM candidate is that it couples mainly to the photon and therefore interacts mostly with protons, and secondly and more importantly it has long-range type interactions modifying substantially the recoil spectrum with respect to a contact interaction. In fact, as pointed out in \cite{Fornengo:2011sz}, low-energy threshold detectors like CoGeNT and DAMA are especially sensitive to this kind of interaction, and therefore the hope exists to get a better agreement among the various direct detection experiments.

Here we will show how the DM magnetic moment interaction affects the interpretation of the different direct detection experiments. In Sec.~\ref{sec:model} we present the DM interaction Lagrangian and provide some considerations concerning the scattering of DM off nuclei. This will allow us to determine the nuclear recoil rate for a given experiment. In Sec.~\ref{sec:theopred} we show how the experimental favored regions and constraints are modified with respect to the na\"ive contact interaction case. In Sec.~\ref{sec:data} we describe the experimental data set we use and our statistical analysis performed to determine the favored regions and constraints for the DM candidate envisioned here. We summarize these results in Sec.~\ref{sec:results}. We determine the associated thermal relic density Sec.~\ref{sec:omega}, and in Sec.~\ref{sec:constraints} we report the constraints imposed by indirect searches, colliders and by the observations of compact stars. Finally, we conclude in Sec.~\ref{sec:conclusions}.

\section{The event rate}
\label{sec:model}

\subsection{Kinematics}

When a DM particle scatters off a nucleus, depending on the DM properties, one can envision at least two distinct kinematics, the elastic and the inelastic. The elastic scattering is represented by
\beq
\chi+ {\cal N}(A,Z)_{\rm at \,rest}\rightarrow \chi+ {\cal N}(A,Z)_{\rm recoil} \ ,
\label{eq:elastic}
\eeq
while the inelastic is
\beq
\chi+ {\cal N}(A,Z)_{\rm at \,rest}\rightarrow \chi'+ {\cal N}(A,Z)_{\rm recoil} \ .
\label{eq:inelastic}
\eeq
In (\ref{eq:elastic}) and (\ref{eq:inelastic}), $\chi$ and $\chi'$ are two DM particle states, and $A$, $Z$ are respectively the mass and atomic numbers of the nucleus 
$\cal N$. In the detector rest frame, a DM particle
with velocity $v$ and mass $m_\chi$ can scatter off a nucleus of mass $m_N$, causing it to recoil. The minimal velocity providing a recoil energy $E_{R}$ is:
\beq
\label{minvelocity}
v_{\rm min}(E_{R}) \simeq \sqrt{ \frac{m_{N}\,E_{R}}{2\mu_{\chi {N}}^2} }\left(1+\frac{\mu_{\chi {N}}\,\delta}{m_{N}\,E_{R}}\right) \ ,
\eeq
where $\mu_{\chi N}$ is the DM-nucleus reduced mass and  $\delta=m_\chi'-m_\chi$ is the mass splitting between $\chi$ and $\chi'$, and the equation above holds for $\delta \ll m_\chi', m_\chi$. Elastic scattering occurs for $\delta = 0$, while $\delta \neq 0$ implies inelastic scattering. In this paper we will only consider the case of elastic scattering.

\subsection{Model and differential cross section}
We concentrate on the possibility of having a massless mediator of the DM-nuclei interaction able to yield long-range interactions absent in contact interactions usually assumed. The obvious candidate mediator in the SM is the photon while other exotic possibilities can be also envisioned such as a dark photon, see e.g.~\cite{Fornengo:2011sz, Chun:2010ve}. 

Assuming that the dark matter is either a neutral boson or fermion one can then use the language of the effective theories to select the relevant operators. Stability of the specific candidate can be easily ensured by charging it under an unbroken global symmetry. The most popular choices are either a new $U(1)$ or a $\mathbb{Z}_2$ depending on the reality of the specific candidate DM field. 

In \cite{DelNobile:2011uf}  one can find a general classification, up to dimension six, of all the possible interactions between a scalar DM and the SM fields. To this order one discovers that there is only one gauge-invariant operator coupling the DM complex boson to a photon, namely
\beq
i (\varphi^\dagger \overleftrightarrow{\partial_\mu} \varphi) \partial_\nu F^{\mu\nu} \ ;
\eeq
DM-nucleus scattering mediated by this operator has been studied for instance in Ref.~\cite{Foadi:2008qv,Frandsen:2009mi,DelNobile:2011je}, where it is shown that it actually leads to a contact interaction and will henceforth not be used here.

On the other hand, considering a fermionic DM $\chi$ one can show that the only 
gauge-invariant couplings to the photon, up to dimension five, are the magnetic and electric moment-mediated interactions:
\beq\label{Lint}
\mathcal{L}^{\rm int}_\text{M} = -\frac{1}{2} \lambda_\chi \, \bar\chi \sigma_{\mu \nu} \chi F^{\mu\nu} \ ,
\qquad\qquad
\mathcal{L}^{\rm int}_\text{E} = -\frac{1}{2} i \, d_\chi \, \bar\chi \sigma_{\mu \nu} \gamma^5 \chi F^{\mu\nu} \ .
\eeq
For a Majorana fermion both $\mathcal{L}^{\rm int}_\text{M}$ and $\mathcal{L}^{\rm int}_\text{E}$ vanish identically, therefore $\chi$ has to be a Dirac fermion. $\lambda_\chi$ and $d_\chi$ are the magnetic and electric dipole moment, respectively, and are usually expressed in units of $e \times \text{cm}$. The scales $\Lambda_\text{M} \equiv e / \lambda_\chi$ and $\Lambda_\text{E} \equiv e / d_\chi$ can be interpreted as the energy scales of the underlying interaction responsible for the associated operators to arise. For instance, in models in which the interactions \eqref{Lint} are explained by the DM being a bound state of charged particles, the $\Lambda$'s could represent the compositeness scale.
 %The inverse of the magnetic and electric dipole moments, respectively $\lambda_\chi$ and $d_\chi$, can be interpreted as the energy scale of the underlying interaction responsible for the associated operators to arise. For instance, in models in which the interactions \eqref{Lint} are explained by the DM being a bound state of charged particles, $\lambda_\chi^{-1}$ and $d_\chi^{-1}$ could represent the compositeness scale.

%We consider %here 
%a DM particle of spin $1/2$ with electric dipole moment $d_\chi$ and magnetic dipole moment $\lambda_\chi$. Specifically, such a particle couples to the electromagnetic field according to the following interaction Lagrangian:
%\beq\label{Lint}
%\mathcal{L}_{\rm int}= -\frac12 \bar\chi \sigma_{\mu \nu} \left(\lambda_\chi+i \, d_\chi\gamma^5 \right) \chi F^{\mu\nu} \ ,
%\eeq
%where we wrote $\chi$ as a Dirac fermion. Due to the coupling with a massless mediator such as the photon, the interaction is of long-range type and therefore, as pointed out in \cite{PFR}, low-energy threshold detectors (like CoGeNT and DAMA) are especially sensitive.

%The differential cross sections for spin-independent scattering\footnote{We neglect the spin-dependent part of the cross section. Since it bears no $Z^2$ enhancement, it is negligible with respect to the spin-independent one.} with nuclei in the non-relativistic limit %\xxx{\cite{xx1, xx2}} %corresponding to the interaction Lagrangians in \eqref{Lint}
The differential cross sections for elastic scattering are~\cite{Barger:2010gv}
\begin{align}
\label{diffCSM}
\begin{split}
&\frac{d\sigma_\text{M}(v,E_R)}{dE_R} = \frac{d\sigma^{\rm SI}_\text{M}(v,E_R)}{dE_R}+\frac{d\sigma^{\rm SD}_\text{M}(v,E_R)}{dE_R}=\\
&\frac{\alpha \lambda_\chi^2}{E_R}  \left\{Z^2 \left[1-\frac{E_R}{v^2}\left(\frac{2m_N+m_\chi}{2m_N m_\chi}\right)\right] F_{\rm SI}^2(E_R) +\left(\frac{\bar \lambda_{\rm nuc}}{\lambda_p}\right)^2 \frac{E_R}{v^2}  \frac{m_N}{3 m_p^2} F_{\rm SD}^2(E_R) \right\}\ ,
\end{split}
\\
\label{diffCSE}
&\frac{d\sigma_\text{E}(v,E_R)}{dE_R} = \frac{\alpha Z^2}{E_R v^2} \, d_\chi^{\, 2} \, F_{\rm SI}^2(E_R) \rule{0pt}{30pt}  \ ,
\end{align}
where $v$ is the speed of the DM particle in the Earth frame, $\alpha=e^2/4\pi\simeq 1/137$ is the fine structure constant, $\lambda_p=e / 2 m_p$ is the nuclear magneton and $F_{\rm SI}$ $(F_{\rm SD})$ denotes the spin-independent (spin-dependent) nuclear form factor which takes into account the finite dimension of the nucleus. Here 
\beq
\bar\lambda_{\rm nuc} = \left(\sum_{\rm {\it i}sotopes}f_i \mu_i^2 \frac{S_i+1}{S_i}\right)^{1 / 2} \ ,
\eeq
is the weighted dipole moment of the target \cite{Chang:2010en}, where $f_i$, $\mu_i$ and $S_i$ are respectively the abundance, nuclear magnetic moment and spin of the $i$-th isotope; the values for these quantities are taken from \cite{PDG, IAEA} and agree exactly with the values provided in Fig.~1 of \cite{Chang:2010en}.

The differential cross section in~\eqref{diffCSM} features both a spin-independent (SI) and a spin-dependent (SD) part. The SI part, neglecting for a moment the form factor, contains two terms: an energy dependent term with a $E_R^{-1}$ drop-off of the cross section, and an energy independent one. In contrast the common contact interactions only feature the constant term, typically with the $1 / v^2$ dependence on the DM velocity; the SD part is of this kind. Notice that, for low enough energies (that might be also below threshold for some experiments, in principle), the interaction is always SI due to the $E_R^{-1}$ divergence. As the recoil energy rises, the interaction becomes mostly SD for target nuclei with large magnetic moment, namely $^{19}$F, $^{23}$Na and $^{127}$I, ($\bar\lambda_{\rm nuc} / \lambda_p = 4.55,\, 2.86,\, 3.33$ respectively); the SD term is instead negligible for all the other nuclei.

%. For elastic magnetic moment interaction the spin-dependent part turns out to be important for nuclei with large magnetic moment $\lambda_N$, such as F$_{19}$, Na$_{23}$ and , which are the  since it bears no $Z^2$ enhancement (see the appendix where the reader can find more details). This is not the case for inelastic magnetic moment interaction, since the spin-independent differential cross section contains a new negative term that, for large enough $\delta$, can reduce sizeably the differential cross section~\cite{Chang:2010en}.}

%Despite the high magnetic moment of Iodine nuclei, heavy elements feature an even smaller value of $r$; the fact that the SD DM-Iodine scattering cross section plays an important role in \hhref{1007.4200} is therefore to ascribe exclusively to the dynamics of the inelastic scattering.

%Neglecting for a moment the form factor, the differential cross section in (\ref{diffCSM}) has two terms: an energy dependent term with a $E_R^{-1}$ drop-off of the cross-section, and an energy independent one. In contrast the common contact interactions only feature the constant terms, typically with the $1 / v^2$ dependence on the DM velocity. 

The differential cross section in \eqref{diffCSE}, instead, has only an energy dependent term. We observe that  the cross section in the electric case is enhanced with respect to the magnetic one by a factor $1/v^2 \sim 10^6$ translating in a value for $d_\chi$ circa $10^3$ times lower than the one for $\lambda_\chi$, when one tries to fit the experiments. 
This is confirmed for instance by Ref.~\cite{Barger:2010gv}, where the authors find that, in order to fit the CoGeNT data alone, a $\Lambda_\text{M}$ of the order of the TeV is needed for a magnetic moment interaction, while for electric moment interaction $\Lambda_\text{E}$ is around the PeV. 
%This is confirmed for instance by Ref.~\cite{Barger:2010gv}, where the authors find that, in order to fit the CoGeNT data alone, a new physics scale of the order of the TeV is needed for a magnetic moment interaction, while for electric moment interaction the scale raises up to the PeV. 
Given that such a high scale is hardly reconcilable with other attempts to study DM and in general new physics, we will treat from now on only the magnetic dipole moment interaction $\mathcal{L}^{\rm int}_\text{M}$.

As for $F_\text{SI}$ and $F_\text{SD}$, we use the nuclear form factors provided in Ref.~\cite{Fitzpatrick:2012ix}. For the SI interaction we have checked that $F_{\rm SI} $ matches with the standard Helm form factor \cite{Helm}.
%
%\beq
%\label{NuclFormSI1}
%F(q r_N)=3\,\frac{j_1(q r_N)}{q r_N}\exp[-(q\,s)^2/2] \ ,
%\eeq
%
%where $j_1$ is the spherical Bessel function of the first kind with $n=1$
%\beq
%\label{NuclFormSI2}
%j_1(q r_N)=\frac{\sin (q r_N)}{(q r_N)^2}-\frac{\cos (q r_N)}{(q r_N)} \ .
%\eeq
%
%In Eqs.~\eqref{NuclFormSI1}, \eqref{NuclFormSI2} $r_N$ is the nuclear radius, and $s$ is a measure of the nuclear skin thickness. A good agreement is obtained for $s \simeq (197 \,\mbox{MeV})^{-1}$ and 
%$r_N = ((1/(164 {\rm MeV})A^{1/3})^2-5s^2)^{1/2}$. 
We recall that %this expression of spin-independent form factor is derived assuming a Fermi distribution for the nuclear charge and that  
all the parameters used in the parameterization of the nuclear form factors may be affected by sizable uncertainties.

%\xxx{We will use as free parameters in our analysis the DM magnetic dipole moment $\lambda_\chi$ and the DM mass $m_\chi$, 'sta frase da sistemare da qualche parte}

\subsection{Nuclear recoil rate}\label{sec:nucrecrat}

The differential recoil rate of a detector can be defined as:

\beq\label{GeneralRate}
\frac{dR}{dE_{R}}=N_T \int \frac{d\sigma_\text{M}(v,E_{R})}{dE_{R}}\, v \, dn_\chi \ , 
\eeq
where $N_{T}=N_A/A$ is the total number of targets in the detector ($N_A$ is the Avogadro's number) and $dn_\chi$ is the local number density of DM particles with velocities in the elemental volume $d^3v$ around $\vec v$. This last factor can be expressed as a function of the DM velocity distribution $f_\text{E}(\vec v)$ in the Earth frame, which is related to the DM 
velocity distribution in the galactic frame $f_\text{G}(\vec w)$ by the galilean velocity
transformation $f_\text{E}(\vec v) = f_\text{G}(\vec v + \vec v_\text{E}(t))$; here $\vec v_\text{E} (t)$ is the time-dependent Earth (or detector) velocity with respect to the galactic frame. The prominent
time-dependence (on the time-scale of an experiment)
comes from the annual rotation of the Earth around the Sun, which is the origin of the
annual modulation effect of the direct detection rate \cite{Drukier:1986tm, Freese:1987wu}. More specifically:

\beq\label{vE}
\vec v_\text{E}(t) = \vec v_\text{G} + \vec v_\text{S} + \vec v_\oplus(t) \ .
\eeq
The galactic rotational velocity of our local system $\vec v_\text{G}$ and the Sun's proper motion
$\vec v_\text{S}$ are basically aligned and their absolute values are $v_\text{G} \equiv v_0 = 220 \pm 50$ km/s and $v_\text{S} = 12$ km/s, while the Earth rotational velocity $\vec v_\oplus(t)$ has
a size $v_\oplus = 30$ km/s, period of 1 year and phase such that it is aligned to
$\vec v_\text{G}$ around June 2$^\text{nd}$ and it is inclined of an angle $\gamma \simeq 60^\circ$ with
respect to the galactic plane. More details can be found, for instance, in Ref.~\cite{Fornengo:2003fm}.
Summarizing:
\beq\label{densitynumberDD}
dn_\chi=n_\chi f_{\rm E}(\vec v)\,d^3v \ ,
\eeq
where $n_\chi=\xi_\chi\rho_0/m_\chi$ is the local DM number density in the Galaxy
and is determined by the local Dark Matter density $\rho_0$ and, in general,
by a scaling factor $\xi_\chi$ which accounts for the possibility that
the specific DM candidate under consideration does not represent the whole amount of DM.
Here we assume $\xi_\chi = 1$. In Eq.~(\ref{densitynumberDD}) the velocity distribution function needs to be properly normalized: this
can be achieved by requiring that in the galactic frame

\beq
\int_{v\leqslant v_{\rm esc}} d^3v \,f_\text{G}(\vec v) = 1 \ ,
\eeq
where $v_{\rm esc}$ denotes the escape velocity of DM particles 
in the Milky Way. For definiteness, we will adopt here $v_{\rm esc} = 650$ km/s.

When considering the differential cross section given in equation (\ref{diffCSM}), the rate of nuclear recoils reduces to
\beq\label{DRate}
\frac{dR}{dE_R}(t)= 
N_A \frac{\xi_\chi\, \rho_0}{A\, m_\chi} \frac{ \alpha \lambda_\chi^2}{E_R} \left(Z^2 \mathcal G_{\rm SI}(v_{\rm min},t) F_{\rm SI}^2(E_R) + \left(\bar\lambda_{\rm nuc}/\lambda_p\right)^2 \mathcal G_{\rm SD}(v_{\rm min},t) F_{\rm SD}^2(E_R) \right) \ ,
\eeq
where
\begin{align}
\mathcal G_{\rm SI}(v_{\rm min}(E_R), t) &= \mathcal I(v_{\rm min}(E_R),t) \left[\frac{\mathcal I_1(v_{\rm min}(E_R),t)}{\mathcal I(v_{\rm min}(E_R),t)}-E_R \left(\frac{2m_N+m_\chi}{2m_N m_\chi}\right)\right] \ ,
\\
\mathcal G_{\rm SD}(v_{\rm min}(E_R), t) &= \mathcal I(v_{\rm min}(E_R),t) \frac{m_N E_R}{3m_p^2} \rule{0pt}{20pt} \ ,
\end{align}
and 
\beq
{\mathcal I(v_{\rm min},t)} = \int_{v \geqslant v_{\rm min}(E_R)} \hspace{-.85cm}d^3 v \,\, \frac{f_\text{E}(\vec v)}v \ , \qquad
{\mathcal I_1(v_{\rm min},t)} = \int_{v \geqslant v_{\rm min}(E_R)} \hspace{-.85cm}d^3 v \,\,v \, f_\text{E}(\vec v) \ , 
\eeq
with $v_{\rm min} (E_R)$ given by Eq.~(\ref{minvelocity}). The detection rate is function of time through the velocity integrals $\mathcal I(v_{\rm min},t)$ and $\mathcal I_1(v_{\rm min},t)$
as a consequence of the annual motion of the Earth around the Sun. Their actual form depends on the velocity distribution function of the DM particles in the halo. In this paper we will consider an isothermal sphere density profile for the DM, whose velocity distribution function in the galactic frame is a truncated Maxwell-Boltzmann:
\beq
f_\text{G}({\vec v}) = \frac{\exp(- v^2 / v_0^2)}{(v_0 \sqrt{\pi})^3 \, \text{erf}(v_\text{esc} / v_0) - 2 v_0^3 \pi (v_\text{esc} / v_0) \exp(- v_\text{esc}^2 / v_0^2)} \ .
\eeq
Under this assumption, and defining the normalized velocities $\eta_\odot \equiv (v_\text{G}+v_\text{S})/v_0 \equiv v_\odot/v_0$, $\eta_\text{E}(t) \equiv v_\text{E}(t)/v_0$, $\eta_{\rm min}(E_R) \equiv v_{\rm min}(E_R)/v_0$ and $\eta_{\rm esc} \equiv v_{\rm esc}/v_0$, the velocity integrals can be written analytically as \cite{Barger:2010gv}
\beq\label{velocityI}
\mathcal I(\eta_{\rm min},t) =\frac1{2\,v_0 \eta_\text{E}(t)}\left[\mbox{erf}(\eta_+)-\mbox{erf}(\eta_-)\right] -\frac1{\sqrt\pi \,v_0 \eta_\text{E}(t)}\left(\eta_+-\eta_-\right)e^{-\eta_{\rm esc}^2}
\eeq
and
\beq
\mathcal I_1(\eta_{\rm min},t) =v_0\left[\left(\frac{\eta_-}{2\sqrt\pi\,\eta_\text{E}(t)}+\frac1{\sqrt\pi}\right)e^{-\eta_-^2}-\left(\frac{\eta_+}{2\sqrt\pi\,\eta_\text{E}(t)}-\frac1{\sqrt\pi}\right)e^{-\eta_+^2}\right] \nonumber
\eeq
\beq\label{velocityI1} 
+\frac{v_0}{4\,\eta_\text{E}(t)}\left(1+2\eta_\text{E}^2(t)\right)\left[\mbox{erf}(\eta_+)-\mbox{erf}(\eta_-)\right] \rule{0pt}{20pt}
\eeq
\beq
-\frac{v_0}{\sqrt\pi}\left[2+\frac1{3\eta_\text{E}(t)} \left(\left(\eta_{\rm min}+\eta_{\rm esc}-\eta_-\right)^3-\left(\eta_{\rm min}+\eta_{\rm esc}-\eta_+\right)^3\right)\right]e^{-\eta_{\rm esc}^2} \ , \nonumber
\eeq
%
%
%\begin{multline}\label{velocityI1} 
%\mathcal I_1(\eta_{\rm min},t) =v_0\left[\left(\frac{\eta_-}{2\sqrt\pi\,\eta_\text{E}(t)}+\frac1{\sqrt\pi}\right)e^{-\eta_-^2}-\left(\frac{\eta_+}{2\sqrt\pi\,\eta_\text{E}(t)}-\frac1{\sqrt\pi}\right)e^{-\eta_+^2}\right]
%\\
%+\frac{v_0}{4\,\eta_\text{E}(t)}\left(1+2\eta_\text{E}^2(t)\right)\left[\mbox{erf}(\eta_+)-\mbox{erf}(\eta_-)\right] \rule{0pt}{20pt}
%\\
%-\frac{v_0}{\sqrt\pi}\left[2+\frac1{3\eta_\text{E}(t)} \left(\left(\eta_{\rm min}+\eta_{\rm esc}-\eta_-\right)^3-\left(\eta_{\rm min}+\eta_{\rm esc}-\eta_+\right)^3\right)\right]e^{-\eta_{\rm esc}^2} \ ,
%\end{multline}
%
where $\eta_\pm(E_R)=$ min$( \eta_{\rm min}(E_R)\pm\eta_\text{E},\eta_{\rm esc})$.

Since in Eq.~\eqref{vE} the rotational velocity of the Earth around the Sun $v_\oplus$, is relatively small compared to the main contribution represented by $v_\text{G} + v_\text{S}$, we can approximate $\vec v_\text{E}(t)$ with its component directed toward the galactic center. We can then write \cite{Fornengo:2003fm}
\beq\label{vEarth}
\eta_{\rm E}(t) \simeq \eta_\odot+\Delta\eta \, \cos\left[2\pi(t-\phi)/\tau\right] \ ,
\eeq 
where $\Delta \eta = v_\oplus\cos\gamma/v_0$, with $\Delta\eta \ll \eta_\odot$, and where $\phi = 152.5$ days (June 2$^\text{nd}$) is the phase and $\tau=365$ days is the period of the Earth motion around the Sun. %The case in which this approximation departs maximally from the actual value is (\xxx{should be?}) when $\vec v_\oplus$ is orthogonal to $\vec v_\text{G}$, and therefore in this case the relative error is $1 - (1 + (v_\oplus / v_\odot)^2)^{1/2} \approx 1\%$.
%\\
By means of Eq.~(\ref{vEarth}) we can then expand the recoil rate, assuming that the velocity distribution
is not strongly anisotropic:
\beq
\frac{dR}{dE_R}(t) \simeq \label{firstTaylorRate} \left.\frac{dR}{dE_R}\right|_{\eta_{\rm E}=\eta_\odot}
+\frac\partial{\partial\eta_{\rm E}}\left.\frac{dR}{dE_{R}}\right|_{\eta_{\rm E}=\eta_\odot}
\Delta\eta \cos\left[2\pi(t-\phi)/\tau\right] \ .
\eeq

To properly reproduce the recoil rate measured by the experiments, we should take into account the effect of partial recollection of the released energy (quenching), and the energy resolution of the detector:
\beq
\frac{dR}{dE_{\rm det}}(E_{\rm det})=\int dE'\,\mathcal{K}(E_{\rm det},E')\sum_{i}\frac{dR_i}{dE_{R}}\left(E_{R}=\frac{E'}{q_i}\right).
\label{eq:recoil}
\eeq
Here the index $i$ denotes different nuclear species in the detector,
$E_{\rm det}$ is the detected energy and $q_i$ are the quenching factors for each of the
nuclear species. The function $\mathcal{K}(E_{\rm det},E')$ reproduces the effect of the energy resolution of the detector; as is generally done, we assume for it a Gaussian behavior.

Finally, the recoil rate of Eq.~(\ref{eq:recoil}) must be averaged over the
energy bins of the detector. For each energy bin $k$ of width
$\Delta E_k$ we therefore define the unmodulated components of the
rate $S_{0k}$ and the modulation amplitudes $S_{\text{m}k}$ as:

\beq\label{ExpTotalRateTay1}
S_{0k}=\frac1{\Delta E_k}\int_{\Delta E_k}dE_{\rm det}\,\left.\frac{dR}{dE_{\rm det}}\right|_{\eta_{\rm E}=\eta_\odot} \ ,
\eeq

\beq\label{ExpTotalRateTay2}
S_{\text{m}k}=\frac1{\Delta E_k}\int_{\Delta E_k}dE_{\rm det}\,
\left.
\frac\partial{\partial\eta_{\rm E}}\frac{dR}{dE_{\rm det}}
\right|_{\eta_{\rm E}=\eta_\odot}\Delta\eta \ .
\eeq
$S_{0k}$ and $S_{\text{m}k}$ are the relevant quantities that we use for the analysis of the experiments which address the annual modulation effect, namely DAMA and CoGeNT. For the other experiments, only the $S_{0k}$ are relevant.

\section{Theoretical predictions}
\label{sec:theopred}

\begin{figure}[!t]
\centering
\includegraphics[width=.6\textwidth]{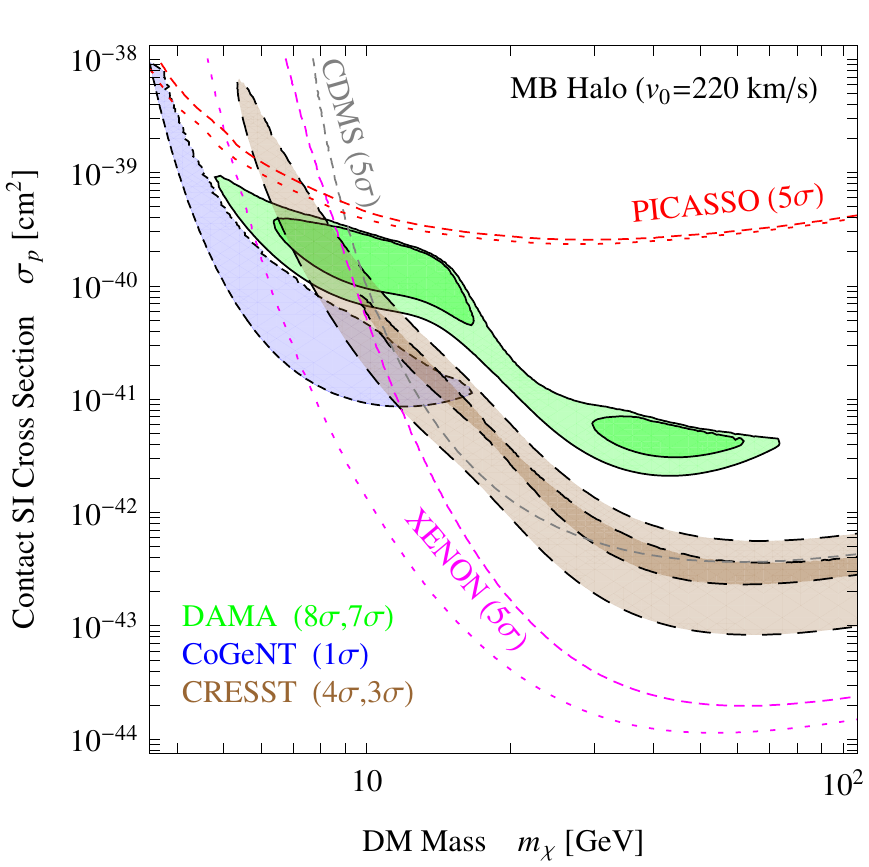}
\caption{DM-proton spin-independent interaction cross section $\sigma_p$ %{\color{red}che va come $\lambda_\chi^2$} 
as a function of the Dark Matter mass $m_\chi$, in the ``standard'' case of coherent contact interaction. The galactic halo has been assumed in the form of an isothermal sphere with velocity dispersion $v_0=220$ km/s and local density $\rho_0 = 0.3$ GeV/cm$^3$. In this figure we show the allowed regions compatible with the annual modulation effects in DAMA and CoGeNT, as well as the region compatible with the CRESST excess, when interpreted as a DM signal. Specifically, the solid green contours denote the regions compatible with the DAMA annual modulation effect \cite{Bernabei:2008yi,Bernabei:2010mq}, in absence of channeling \cite{Bernabei:2007hw}. The short-dashed blue contour refers to the region derived from the CoGeNT annual modulation effect \cite{Aalseth:2011wp}, when the bound from the unmodulated CoGeNT data is included. The dashed brown contours denote the regions compatible with the CRESST excess \cite{Angloher:2011uu}. For all the data sets, the contours refer to regions where the absence of excess can be excluded with a C.L. of $7\sigma$ (outer region), $8\sigma$ (inner region) for DAMA, $1\sigma$ for CoGeNT and $3\sigma$ (outer region), $4\sigma$ (inner region) for CRESST. For XENON, the constraints refer to a threshold of 4 photoelectrons (published value \cite{Aprile:2011hi}, lower line) and 8 photoelectrons (our conservative estimate, upper line), as discussed in Sec.~\ref{sec:data}. The two lines for PICASSO enclose the uncertainty in the energy resolution \cite{Archambault:2012pm}.}
\label{fig:CSI}
\end{figure}
\begin{figure}[!t]
\includegraphics[width=0.49\textwidth]{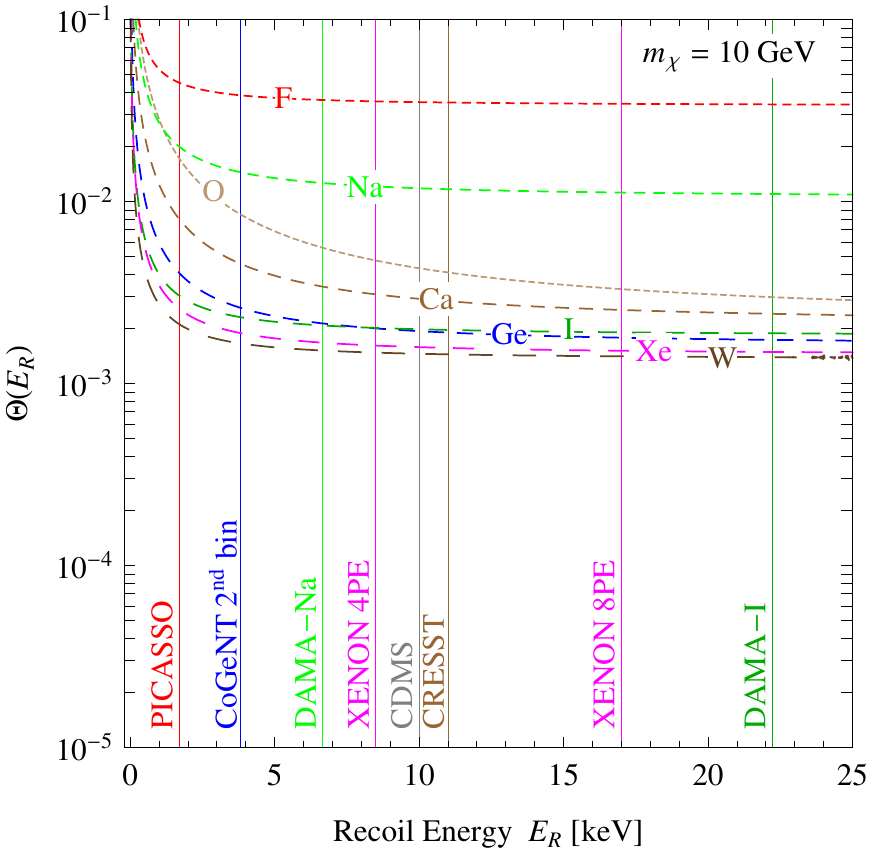}
\includegraphics[width=0.49\textwidth]{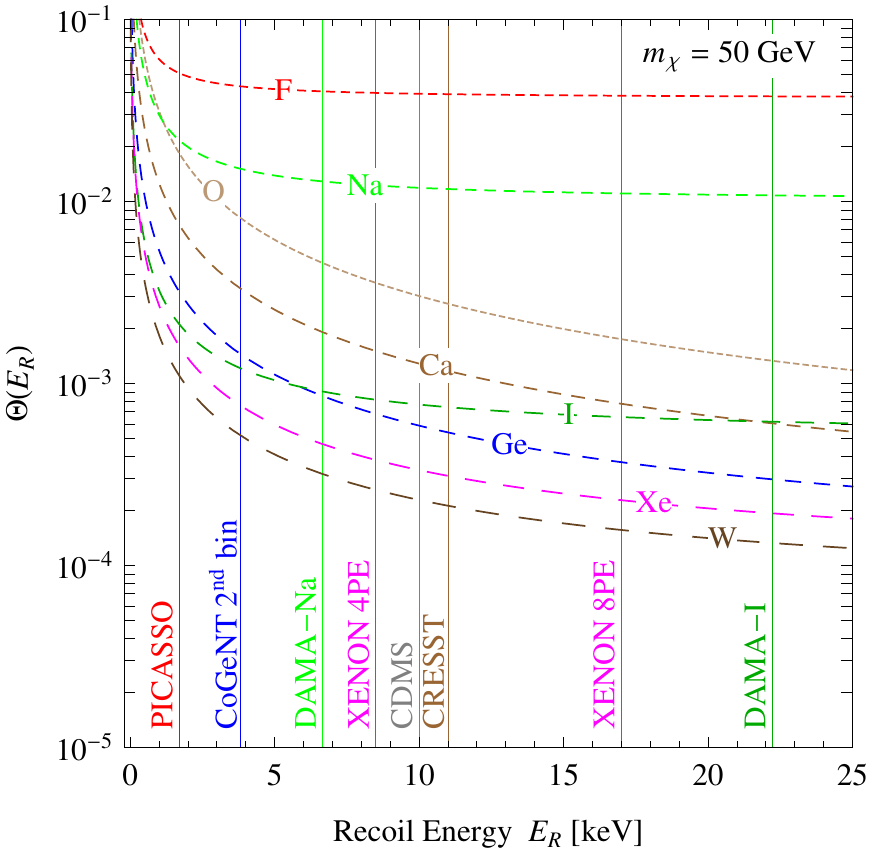}
\caption{The function $\Theta(E_R)$ (dashed lines), parametrizing the deviation of the allowed regions and constraints with respect to the standard case of contact interaction without isospin violation presented in Fig.~\ref{fig:CSI}. Each line is for one of the target elements used in the experiments taken into account in the text. The vertical lines indicate the low energy thresholds for the various experiments (apart from CoGeNT, for which we show the threshold of the second energy bin, where much of the signal is recorded). For DAMA two thresholds are shown, corresponding to the corrections due to the quenching factors for sodium ($q_{\rm Na} = 0.3$, in the assumption of scattering mainly off Na) and iodine ($q_{\rm I} = 0.09$, for the assumption of scattering mainly off I). For XENON the 4 photoelectrons (4PE, published value \cite{Aprile:2011hi}) and 8 photoelectrons (8PE, our conservative estimate) thresholds are shown, as discussed in Sec.~\ref{sec:data}. We have shown also the lowest threshold for the PICASSO experiment, although we refer to the main text for a more precise interpretation of its temperature dependent thresholds. The picture is reported for two values of the DM mass, $m_\chi = 10$ GeV (left panel) and $m_\chi = 50$ GeV (right panel).  
}
\label{fig:A}
\end{figure}

We attempt here an analytical study of the effect of the magnetic moment interaction on the observed differential rate. More precisely we provide a simple comparison between the results arising from the interaction studied here and the {\it standard} picture, i.e. the spin-independent coherent contact interaction, shown in Fig.~\ref{fig:CSI}.
%of how the allowed regions and constraints for spin-independent scattering will be modified compare to the standard picture. %Let us now to express the nuclear recoils rate in a more familiar way in order to have a rough interpretation of how the allowed regions and constraints for spin-independent scattering will be modified due to a magnetic moment interaction. 
To this aim, Eq.~(\ref{DRate}) can be rewritten as
\beq
\frac{dR}{dE_R}(t)=N_A \frac{\xi_\chi\, \rho_0}{A\, m_\chi} \frac{m_N}{2 \mu_{\chi n}^2} A^2 \alpha\lambda_\chi^2\,\Theta(E_R) \mathcal I(\eta_{\rm min}(E_R),t) F_{\rm SI}^2(E_R) \ ,
\eeq
where $\mu_{\chi n}$ is the DM-nucleon reduced mass, $\alpha\lambda_\chi^2$ plays the role of the spin-independent DM-proton cross section $\sigma_p$ and we defined%the function
\beq\label{Theta}
\Theta(E_R)= \Theta_{\rm SI}(E_R)\left(1+\frac{\Theta_{\rm SD}(E_R)}{\Theta_{\rm SI}(E_R)}\frac{F_{\rm SD}^2(E_R)}{F_{\rm SI}^2(E_R)}\right)=\Theta_{\rm SI}(E_R)\left(1+r(E_R)\right)\ ,
\eeq
where
\begin{align}
\label{ThetaSI}
& \Theta_{\rm SI}(E_R) &=& \left(\frac ZA\right)^2 \left(\frac{2\mu_{\chi n}^2}{m_N E_R}\right) \left(\frac{\mathcal G_{\rm SI}(\eta_{\rm min}(E_R),t)}{\mathcal I(\eta_{\rm min}(E_R),t)}\right),  
\\
\label{ThetaSD}
& \Theta_{\rm SD}(E_R) &=& \left(\frac1A\right)^2 \left(\frac{\bar \lambda_{\rm nuc}}{\lambda_p}\right)^2 \left(\frac{2\mu_{\chi n}^2}{m_N E_R}\right) \left(\frac{\mathcal G_{\rm SD}(\eta_{\rm min}(E_R),t)}{\mathcal I(\eta_{\rm min}(E_R),t)}\right)=\frac23\left(\frac1A\right)^2\left(\frac{\bar \lambda_{\rm nuc}}{\lambda_p}\right)^2\left(\frac{\mu_{\chi n}}{m_p}\right)^2 \rule{0pt}{25pt} \ .
\end{align}
The function $\Theta$ measures the deviation of the allowed regions and constraints with respect to the standard spin-independent picture, and the ratio $r$ parametrizes the relevance of the SD interaction respect to the SI one: if $r>1$ the interaction is largely SD, while it is mostly SI for $r<1$ %. We have found that the dipole-dipole term plays a major role for DAMA ($^{23}$Na and $^{127}$I) and PICASSO ($^{19}$F), while being negligible for all the other experiments considered in our analysis 
(see Appendix~\ref{appendix} for a more detailed analysis). %in the plane $(m_\chi, \sigma_p)$ because it embeds all of the residual dependence on the features of different detectors ($m_N$, $A$, $Z$ and different energy thresholds). Furthermore, 

Since $\mathcal G_{\rm SI}/\mathcal I$ depends itself on $m_\chi$, we expect an asymmetric shift of the favored regions and constraint lines in the $(m_\chi, \sigma_p)$ plane due exclusively to the modified dynamics of the interaction considered here. %in the normalization moving from light to heavy DM mass is expected.
%This rescaling of the DM parameter space is very predictive because $\Theta$ does not involve any further free parameters.
There are only two free parameters in this model: $m_{\chi}$, and $\sigma_p$ (or alternatively $\lambda_{\chi}$).
 This is not true for instance in models with contact interaction in which the DM interacts differently with protons and with neutrons (the so-called ``isospin violating'' scenario, see e.g.~\cite{Gondolo:2011eq,DelNobile:2011je,Feng:2011vu,Chun:2010ve,DelNobile:2011yb,Gao:2011ka,Pato:2011de,Chen:2011vd,Frandsen:2011cg,Kurylov:2003ra,Cline:2011zr,Cline:2011uu}). Indeed in such models the role of $\Theta$ is played by the expression $\left([Z+f_n/f_p(A-Z)]/A\right)^2$, and therefore there are three free parameters that must be fitted: $\sigma_p$, $m_\chi$, and the ratio of the DM-neutron and DM-proton couplings $f_n/f_p$. 

 In Fig.~\ref{fig:A} we show the behavior of $\Theta$ as a function of $E_R$ both for small (left panel) and large (right panel) DM masses, considering several targets, $v_0=220$ km/s and $v_\text{E}=v_\odot$ (being $v_\oplus\cos\gamma \ll v_\odot$). In order to simplify the reading of the figures, we take the limit $v_\text{esc} \rightarrow \infty$; one can figure out the effect of a finite escape velocity as a target-dependent cut-off at high recoil energies, that sets in when the minumum velocity required to have a recoil above threshold exceeds $v_\text{esc} + v_\text{E}$, the maximum escape velocity in the Earth frame. There, the function $\Theta$ loses its meaning since the expected rate is zero both in the standard picture and in the case considered here.

A first striking feature of $\Theta$ is its overall magnitude: depending on the DM mass and on the experiment, we have in fact a suppression of the expected rate of roughly 1 to 4 orders of magnitude. For nuclei in which the interaction is mostly SD ($r>1$), $\Theta$ provides a suppression of 1 to 3 orders of magnitude for the well known reason that the rate does not carry any $A^2$ factor, contrarily to the usual SI case. Due to this fact, the SD part of the cross section is usually considered negligible respect to the SI one, but this turns out to be false in our case, for nuclei with large magnetic moment, since also the SI part is strongly suppressed. %While the last term in Eq.~\eqref{ThetaSD} is close to unity for $m_\chi\gg m_p$, the other factors give a suppression of $\Theta$ of roughly 1 to 3 orders of magnitude.
This suppression is due to the interplay of the last two factors in Eq.~\eqref{ThetaSI}, while the first term plays only a marginal role given that $A \sim 2 Z$ roughly for all the target nuclei. The second term, $2\mu_{\chi n}^2 / m_N E_R$, enhances $\Theta$ by 2 to 4 orders of magnitude due to the presence of $E_R$, whose typical scale is few (tens of) keV and therefore very small compared to all the other mass scales involved. The last term $\mathcal{G}_\text{SI} / \mathcal{I}$, containing the velocity integrals, gives instead a suppression of roughly 6 orders of magnitude.
Given the measured rate, this overall suppression provided by the $\Theta$ function will reflect in the fit pointing to higher interaction cross sections, with respect to the standard case of Fig.~\ref{fig:CSI}. This fact will play a role in fitting the relic abundance (Sec.~\ref{sec:omega}).

The steep rise at low energies is always due to the long-range type of the SI part of the interaction, being the SD one of a contact type; in the particular case of magnetic dipole moment interaction considered here, the differential cross section diverges as $1 / E_R$. At energies higher than the steep rise, for nuclei with large magnetic moment ($r>1$), the interaction rapidly becomes of a contact type and therefore the function $\Theta$ exhibits a plateau whose value is  $\sim (\bar\lambda_{\rm nuc}/A \, \lambda_p)^2$. Instead, for the other nuclei for which $r<1$, two regimes are possible, depending on the ratio $v_\text{min} / v_\text{E}$. These reflect the two trends assumed by the function\footnote{Here we ignore the time dependence, which gives only a negligible contribution.} \mbox{$\zeta(v_\text{min}) \equiv \mathcal{I}_1(v_\text{min}) / \mathcal{I}(v_\text{min})$}, namely a plateau $\zeta \sim 1.8 \, v_0^2$ for $v_\text{min} \ll v_\text{E}$ and a rise $\zeta \sim v_\text{min}^2$ for $v_\text{min} \gg v_\text{E}$, with an intermediate value of $\zeta(v_\text{E}) \sim 2.8 \, v_0^2$.

Light DM particles require a higher minimum velocity to recoil compared to heavier ones. For the range of DM masses we are interested in, and for the range of recoil energies relevant for the direct DM search experiments, 
 the value of $v_\text{min} / v_\text{E}$ is  controlled by the ratio $m_N / m_\chi$. We will see that, depending on this ratio, the function $\Theta_{\rm SI}$ can assume a constant behavior or scale as $E_R^{-1}$, while the function $\Theta_{\rm SD}$ is always constant.

\subsection{Light Dark Matter}

For DM particles much lighter than the target nuclei, we have that $v_\text{min} > v_\text{E}$ already at low energies for all the targets considered, and therefore
\beq\label{regime1}
\frac{\mathcal G_{\rm SI}(\eta_{\rm min}(E_R),t)}{\mathcal I(\eta_{\rm min}(E_R),t)}\simeq \left[v_{\rm min}^2(E_R)-E_R \left(\frac{2m_N+m_\chi}{2m_N m_\chi}\right)\right] = \frac{m_N E_R}{2 m_\chi^2} \ ;
\eeq
accordingly, $\Theta$ simplifies to $\left(\mu_{\chi n} / (A \,m_p)\right)^2\left[Z^2\left(m_p/m_\chi\right)^2+2/3\left(\bar\lambda_{\rm nuc}/\lambda_p \right)^2\right]$, displaying the plateau shown in the left panel of Fig.~\ref{fig:A}. As expected, the plateau sets in earlier for heavy targets if the SI interaction dominates ($Z^2\left(m_p/m_\chi\right)^2\gg 2/3\left(\bar\lambda_{\rm nuc}/\lambda_p \right)^2$), while its starting point is independent on the target mass for SD interactions. %On the other hand whether the SD part of the interaction plays a major role the plateau turns out to be independent on the mass of the target. %, for which a faster DM particles is needed to generate a scattering
We notice that, for high enough recoil energies, $\Theta$ always gets to this constant behavior; it is interesting though that it exists a regime in which this happens at the low energies relevant for DM direct detection experiments. An indicative  $m_\chi\sim10$ GeV falls into this case, and as we will show in Sec.~\ref{sec:results} can fit the regions of the parameter space allowed by CoGeNT, DAMA-Na, CRESST-O and CRESST-Ca\footnote{DAMA and CRESST are multi-target detectors and allowed regions at large DM mass correspond to scattering on I for DAMA and W for CRESST, while at small DM mass regions correspond to scattering on Na for DAMA and both O and Ca for CRESST.}.
\\
For an experiment with a lower threshold above the steep rise, the differential rate does not feature any dependence on $E_R$ and it is therefore similar to the rate of spin-independent contact interactions, apart from a different functional dependence on $m_\chi$. % As explained above, although the total rate decreases as a result of the new interaction type, as explained above, the shape of the spectrum is not modified. %Notice that also the dependence on the DM mass is changed with respect to the standard case.
Notice that due to this different dependence on the mass of the DM, the rate is more suppressed for heavier DM particles, and therefore we expect the favored regions in the ($m_{\chi}, \sigma_p$) plane as well as the exclusion lines to raise and slightly tilt at higher $m_\chi$ (compared to the spin-independent contact interaction). 

Considering now that most of the signal in a given detector comes generally from the first energy bin\footnote{In CoGeNT, the larger part of the signal comes instead from the second energy bin.}, i.e.~close to the low energy threshold, we can roughly estimate, before doing any statistical analysis of the data, the shift of the allowed regions and constraints compared to the standard picture. For example, the expected rate of DM scatterings off Na nuclei is reduced by a factor $\sim   10^{-2}$ with respect to the standard case (see Fig.~\ref{fig:A}), and therefore the DAMA favored region, when assumed that most of the signal comes from DM scattering on Na, is shifted up by $\sim  10^2$ in the $(m_\chi, \sigma_p)$ plane. Taking now the DAMA-Na allowed region as benchmark, we can see how the other favorite regions and exclusion lines move with respect to it. CoGeNT moves a factor $\sim 5$ up, making the agreement with DAMA almost perfect. Concerning CRESST, the fit at small DM masses is due equally to O and Ca \cite{Angloher:2011uu} in the standard scenario, from which we conclude that the cross section for DM magnetic dipole interaction with Ca is $\sim 1.5$ times bigger than that with O; the overall effect is an increase in $\sigma_p$, that lifts the favored region to better agree with DAMA and CoGeNT. In detail we expect that the CRESST allowed region moves a factor $\sim 4$ towards the DAMA-Na ballpark. 
Finally, CDMS and XENON move up by a factor $\sim 6.5$ and $\sim 8.5$ respectively, roughly irrespective of the choice of the threshold. This improves once again the agreement with the other experiments. For PICASSO the situation is different due to the special experimental setup in which the low energy threshold is a function of the  temperature, and therefore one should be careful especially if the differential cross section is energy dependent. However, since the scattering in the PICASSO experiment is dominated by the dipole-dipole interaction which is of a contact type, a simple rescaling  is again applicable, and we expect that the constraints will become a factor 2.5 more stringent (see left-panel of Fig.~\ref{fig:A}), compared to the standard picture.
 
\subsection{Heavy Dark Matter}

For higher DM masses, where for instance one can find the regions of the parameter space allowed by DAMA-I and CRESST-W, the effect of the long-range interaction turns to be very evident. As the minimal velocity for  a given nuclear recoil $v_{\rm min}$ becomes smaller than $v_\text{E}$,  the function $\zeta(v_\text{min})$ changes behavior. For $v_\text{min} \ll v_\text{E}$ (i.e.~$m_\chi \gg m_N$)\footnote{This is generally true provided we consider a  recoil energy $E_R$ within our interest: from a minimum threshold value of a few keV up to $25-30$ keV.} 
\beq\label{regime2}
\frac{\mathcal G_{\rm SI}(\eta_{\rm min}(E_R),t)}{\mathcal I(\eta_{\rm min}(E_R),t)}\simeq \left[1.8 \, v_0^2 - E_R \left(\frac{2m_N+m_\chi}{2m_N m_\chi}\right)\right] \ .
\eeq
In this regime, for the nuclei whose interaction is SI, $\Theta$ scales as $E_R^{-1}$ for the whole range of the recoil energies relevant for the direct detection experiments. If $E_R$ increases sufficiently enough, $v_\text{min}$ becomes eventually  larger than $v_\text{E}$ and $\Theta$ assumes the constant behavior we described in the previous section. The difference in $\Theta$ between the low and high DM masses we depicted in Fig.~(\ref{fig:A}) is that in the first case the plateau sets in at small recoil energies (smaller than the energy threshold of most experiments), whereas in the second case, the plateau sets in at large recoil energies (more than 25 keV). Instead for the other nuclei with large magnetic moment, the interaction is SD and as in the case of light DM the plateau manifests itself at low recoil energy.    

Due to the pronounced energy dependence in this case, the allowed regions and the constraints will shift in the ($m_{\chi}, \sigma_p$) plane considerably more  than in the case of low DM mass (still compared to the standard contact spin-independent interaction).  By taking again the DAMA-Na allowed region as a benchmark from the right panel of Fig.~(\ref{fig:A}), we estimate that both DAMA-I and CRESST-W shift more than one order of magnitude up in $\sigma_p$. The same happens for CDMS and XENON. On the other hand PICASSO, like DAMA-Na, does not change its behavior respect to the light DM case: this is due to the fact that $^{19}$F and $^{23}$Na enjoy mostly spin-dependent interactions, that depend only slightly on the DM mass.
\bigskip

To summarize, taking into account the whole DM mass range, we see that, apart from an overall shift upwards in $\sigma_p$, there is a general trend of the various experiments to ``gather'', getting closer to each other, with respect to the standard case. This behavior is not homogeneous in the DM mass and is more pronounced for  masses above $\sim 25$ GeV. The effect is to favor a better fit compared to the standard case, with a better agreement of all the experiments. Even though this is true in general for any DM mass, we don't expect anyway to get a good agreement for higher masses: in that region in fact the XENON experiment rules out the signal featured by other experiments by several orders of magnitude in the standard picture (Fig.~\ref{fig:CSI}), and the shifts that take place in our case are not big enough to change this situation.

Finally we comment on the dependence of the function $\Theta$ from the halo model considered, and in particular from the choice of the dispersion velocity $v_0$. As we have seen, the transition between the two regimes discussed above is driven by the ratio $v_\text{min} / v_\text{E}$, and it is only important for the nuclei which experience SI interactions with a DM particle.  Therefore in this case we expect that for larger values of $v_\text{E} \approx v_0$, the plateau behavior of $\Theta$ sets in at higher recoil energies (or equivalently at smaller DM masses). This would make the shifts more pronounced even for light DM. In Sec.~\ref{sec:results} and  Figs.~\ref{fig:B} and \ref{fig:B-more} we present the exact numerical results.

\section{Data sets and analysis technique}
\label{sec:data}

In this section we discuss  the techniques used to analyze the various
data sets. In particular we adopt the approach of \cite{Fornengo:2011sz}, and we summarize below 
the details of how we perform the fits to data and constraints from null results experiments.

For DAMA, CoGeNT and CRESST, we test the null hypothesis (absence of signal on top of estimated background for CRESST and absence of modulation for DAMA and CoGeNT). From this we infer:

\begin{itemize}
\item[i)] the confidence level for the rejection of the
null hypothesis (we find $8 \div 9 \sigma$ for DAMA,
$1 \div 2 \sigma$ for CoGeNT, and $4 \sigma$ for CRESST);
\item[ii)] the domains in the relevant DM parameter space (defined by the DM mass $m_\chi$ and the DM magnetic dipole moment $\lambda_\chi$) where
the values of the likelihood function depart for more than $n \sigma$ from the
null hypothesis, and thus the corresponding evidence of the DM signal. We use $n=7,8$, $n=1$, and $n=3,4$ for DAMA, CoGeNT, and CRESST, respectively \cite{Belli:2011kw}.

%This choice (test of the null hypothesis) allows more proper comparison between the results arising from experimental data sets with different statistical significances and, for the case of DAMA, allows to implement a requirement of a very high C.L. 
\end{itemize}

Our statistical estimator is the likelihood function of detecting the observed number of events
$\mathcal{L}=\prod_i \mathcal{L}_i$, where the index $i$ indicates the $i$-th energy bin in DAMA and CoGeNT, and the $i$-th detector in CRESST. For DAMA and CoGeNT $\mathcal{L}_i$ are taken to follow a Gaussian distribution and for CRESST,  since in this case the number of events in each sub-detector is low, a Poissonian one.
Defining $\mathcal{L}_\text{bg}$ as the likelihood of absence of signal, we assume the function {$\tilde y=-2\ln{\mathcal{L}_\text{bg}/\mathcal{L}}$} to be distributed as a $\chi^2$-variable with one degree of freedom for a given value of the DM mass (notice that for DAMA and CoGeNT $\tilde y$ reduces to $\tilde y=\chi^2_\text{bg}-\chi^2$). From the $\tilde y$ function, the
interval on $\lambda_\chi$ where the null hypothesis (i.e.~$\lambda_\chi = 0$)
can be excluded at the chosen level of confidence are extracted: $7\sigma$ (outer region) or
$8\sigma$ (inner region) for DAMA, $1 \sigma$
for CoGeNT and $3 \sigma$ (outer region), $4 \sigma$ (inner region) for CRESST. We then plot allowed regions in the $(m_\chi, \lambda_\chi)$ plane.

We derive constraints from CDMS and XENON100 with a similar likelihood function $\lambda=-2\ln{\mathcal{L}/\mathcal{L}_\text{bg}}$; here $\mathcal{L}$ is the likelihood of detecting the observed number of events (2 for CDMS and 3 for XENON100), while in $\mathcal{L}_\text{bg}$ the DM signal is not included. Both likelihoods are taken as Poissonian variables and $\lambda$ is assumed to follow a $\chi^2$-distribution. For PICASSO we instead derive constraints by using a $\Delta \chi^2$ method of the data points shown in fig.~5 of \cite{{Archambault:2012pm}}. Bounds are conservatively shown at 5$\sigma$ C.L.

Concerning the experimental data sets and statistical methods, we refer to \cite{Fornengo:2011sz}; for completeness we summarize below for each experiment the most important ingredients used in the analysis.

\begin{itemize}

\item[$\diamond$] {\bf DAMA}: We use the entire set of DAMA/NaI \cite{Bernabei:2008yi} and DAMA/LIBRA \cite{Bernabei:2010mq} data, corresponding to a cumulative exposure of 1.17 ton$\times$yr.
We analyze the modulation amplitudes $S^{\rm exp}_{\text{m}k}$ reported 
in Fig.~6 of Ref.~\cite{Bernabei:2010mq} requiring that the DM contribution to the unmodulated component of the rate, $S_{0}$, does not exceed the corresponding experimental value $S^{\rm exp}_{0}$ in the $2 \div 4$ keV energy range. We compute
\beq
 \label{eq:yDAMA}
{y = -2\ln{\mathcal{L}}} ~\equiv ~\chi^2(\lambda_\chi,m_\chi) = \sum_{k=1}^8\frac{\left(S_{\text{m}k} - S_{\text{m}k}^{\rm exp}\right)^2}{\sigma_k^2} + 
 \frac{\left(S_{0} - S_{0}^{\rm exp}\right)^2}{\sigma^2} Ê{\theta(S_{0} -S^{\rm exp}_{0} Ê)}\ ,
\eeq
where $\sigma_k$ and $\sigma$ are the experimental errors on $S_{\text{m}k}^{\rm exp}$ and 
$S^{\rm exp}_{0}$, respectively. The last term in Eq.~(\ref{eq:yDAMA}) implements the upper bound on $S_0$ by penalizing the likelihood when $S_0$ exceeds $S^{\rm exp}_0$ with the Heaviside function $\theta$. The detector energy resolution is parametrized by a Gaussian function of width $\sigma_{\rm res}(E)=E(0.448/\sqrt{E}+0.0091)$ \cite{Bernabei:2008yh}, using for the quenching factor the central values quoted by the collaboration, namely $q_{\rm Na} = 0.3$ and $q_{\rm I} = 0.09$ \cite{Bernabei:1996vj}. We don't take into account the possibility of a nonzero channeling fraction \cite{Bernabei:2007hw}.

\item[$\diamond$] {\bf CoGeNT}: We consider the time-series of the data, treating the measured total rate as a constraint. Similarly to the analysis done on the DAMA data, we define
\begin{equation}
 \label{eq:ycogent}
 \begin{split}
& {y = -2\ln{\mathcal{L}} }~\equiv~ \chi^2(\lambda_\chi,m_\chi) = \\
& \sum_{k=1}^{16}\frac{\left(\tilde S_{\text{m}1,k} - \tilde S_{\text{m}1,k}^{\rm exp}\right)^2}{\sigma_k^2} +
\sum_{k=1}^{16}\frac{\left(\tilde S_{\text{m}2,k} - \tilde S_{\text{m}2,k}^{\rm exp}\right)^2}{\sigma_k^2} + \sum_{j=1}^{31}\frac{\left(S_{0j} - S_{0j}^{\rm exp}\right)^2}{\sigma_j^2} {\theta( S_{0j} -S^{\rm exp}_{0j} ) } \ ;
 \end{split}
\end{equation}
here $\tilde S_{\text{m}k}=1/\Delta t_k \int_{\Delta t_k} S_{\text{m}k} \cos\left[2\pi(t-\phi)/\tau\right] dt$, with $\Delta t_k$ the temporal bin of the data, and $\tilde S^\text{exp}_{\text{m}k} = R^\text{exp}_{\text{m}k} - \langle R^\text{exp}_{\text{m}k} \rangle$, where $R^\text{exp}_{\text{m}k}$ is the total rate (taken from Fig.~4 of Ref.~\cite{Aalseth:2011wp}) and $\langle R^\text{exp}_{\text{m}k} \rangle$ is its annual average. The subscripts 1 and 2 in Eq.~\eqref{eq:ycogent} refer to the first and second energy bins. The total rate in the $0.9-3.0$ keV$_{\rm ee}$ energy-bin is computed by subtracting the rate in the $0.5-0.9$ keV$_{\rm ee}$ bin to the rate in the $0.5-3.0$ keV$_{\rm ee}$ bin, with a Gaussian propagation of the errors. $S^{\rm exp}_{0j}$ and $\sigma_j$ denote the experimental counts and the corresponding errors as given in Ref.~\cite{Aalseth:2011wp} (31 energy bins in the interval $0.4 - 2$ keV$_{\rm ee}$), after removal of the L-shell peaks but without removing any other background. The total fiducial mass is 330 g, the energy resolution is given by a Gaussian with width taken from \cite{Aalseth:2008rx}, and the quenching factor below 10 keV is described by the relation $E = 0.2 \, E_R^{1.12}$ \cite{Barbeau:2007qi}.

\item[$\diamond$] {\bf CRESST}: We compute the expected DM signal in each of the 8 CRESST detector modules. The acceptance regions and the number of observed events are provided in Table 1 of Ref.~\cite{Angloher:2011uu}, and we derive background events according to estimates in Sec.~4 of Ref.~\cite{Angloher:2011uu}. A likelihood-ratio test yields a 4.1$\sigma$ C.L. evidence for the best-fit of a DM signal over the background-only hypothesis, in good agreement with the result quoted by the collaboration. We use the published value of 730 kg$\times$days for the exposure and assume an even contribution among the different modules\footnote{This is the same analysis performed in \cite{Fornengo:2011sz}, although a value of 400/9 kg days for the exposure was erroneously reported in the text of the published version.} (each module accounts therefore for an exposure of 730/8 kg$\times$days); we consider moreover a constant efficiency.

\item[$\diamond$] {\bf CDMS}: We use the ``standard'' 2009 CDMS-II results based on Ge data \cite{Ahmed:2009zw}; these are obtained employing conservative nuclear recoil selection cuts and assuming an energy threshold of 10 keV. The total exposure is 612 kg$\times$days and we take the efficiency from the black curve of Fig.~5 in Ref.~\cite{Ahmed:2010hw} with $q \simeq 1$ as quenching factor.\footnote{In the case of light DM, one can perform a similar analysis by using combined data from the CDMS and EDELWEISS experiments (see Fig.~1 in Ref.~\cite{Ahmed:2011gh}), with basically the same results.}
In spite of an expected background of $0.9\pm 0.2$ events, two signal events were found in the $10-100$ keV energy interval \cite{Ahmed:2009zw} (we use these numbers to derive the constraints).

\item[$\diamond$] {\bf XENON100}: We use the results presented in Ref.~\cite{Aprile:2011hi}, with an exposure of 100.9 days in a fiducial volume of 48 kg. After all the cuts, three events were reported in the DM signal region in spite of an expected background of $1.8 \pm 0.6$ events.
We model the data using a Poissonian distribution of photoelectrons, with a single-photoelectron resolution equal to 0.5. The shape of the $\mathcal{L}_{\rm eff}$ function is very crucial for such a low number of photoelectrons and for small DM masses. Following \cite{Fornengo:2011sz} we try to enclose a possible (but not exhaustive%\footnote{For more discussion and considerations about the reliability of the XENON100 constraints for light DM particles see e.g.~\cite{Bernabei:2008jm,Collar:2011wq}.}
) uncertainty on the bounds derived for XENON100 by adopting two different approaches:
\begin{itemize}
\item[i)] we adopt as threshold the published value of 4 photoelectrons and the nominal central value of $\mathcal{L}_{\rm eff}$
as shown in Fig.~1 of Ref.~\cite{Aprile:2011hi}, which relies heavily on linear extrapolation below 3 keV$_\text{nr}$;
\item[ii)] more conservatively, we raise the threshold for the photomultipliers to 8 photoelectrons: this value is the lowest one for which the analysis is nearly independent on the shape of $\mathcal{L}_{\rm eff}$ below 3 keV$_\text{nr}$.
\end{itemize}
Notice that these two approaches are not exhaustive of all the possible assumptions one can do to determine the XENON100 response to light DM (for further discussion and considerations, see e.g.~Refs.~\cite{Bernabei:2008jm,Collar:2011wq}). It appears therefore still preliminary, given these large uncertainties, to assume the bounds quoted by the collaboration as strictly firm. %We nevertheless show them, but without enforcing them in our discussion. In any case, 
The 8 photoelectrons bound is less dependent on the $\mathcal{L}_{\rm eff}$ extrapolation, and therefore, conservatively, we consider it as more appropriate.

Finally, we follow Eqs.~(13--16) in Ref.~\cite{Aprile:2011hx} to compute the expected signal. We derive upper bounds for both CDMS and XENON as mentioned at the beginning of this section.

\item[$\diamond$] {\bf PICASSO}: The PICASSO experiment, located at SNOLAB \cite{Archambault:2012pm}, is very  different from the ones discussed above; it is in fact based on the superheated droplet technique, a variant of the bubble chamber technique, to search for DM recoiling on $^{19}$F in a C$_4$F$_{10}$ target. The experimental procedure consists in measuring the acoustic signal released by the nucleation of a bubble as a function of the temperature $T$.  Details of the detector principle can be found in~\cite{PICASSO1,PICASSO2}. Since bubble formation is only triggered  above a certain  energy threshold $E_{\rm th}(T)$, the spectrum of the particle-induced energy depositions can be constructed by varying the temperature. We compute the predicted DM rate as a function of $E_{\rm th}(T)$ from Eq.~(3) of \cite{Archambault:2012pm} and we compare such prediction with the experimental rate shown in Fig.~5  by using a $\Delta \chi^2$ method. In this analysis we adopt two reference values of the parameter $a(T)$ which describes the steepness of the energy threshold; namely we take $a=(2.5, \ 7.5)$ in order to encapsulate as much as possible the experimental uncertainties \cite{Archambault:2012pm}. We checked our result against the one of the collaboration given in Fig.~7 of Ref.~\cite{Archambault:2012pm}, and we found an excellent agreement. Remaining coherent with our choice of being conservative we show here the result at $5\sigma$, as we did also for the other exclusion experiments.

\end{itemize}

\section{Fit to the direct detection experiments}
\label{sec:results}

\begin{figure}[!t]
\centering
\includegraphics[width=.6\textwidth]{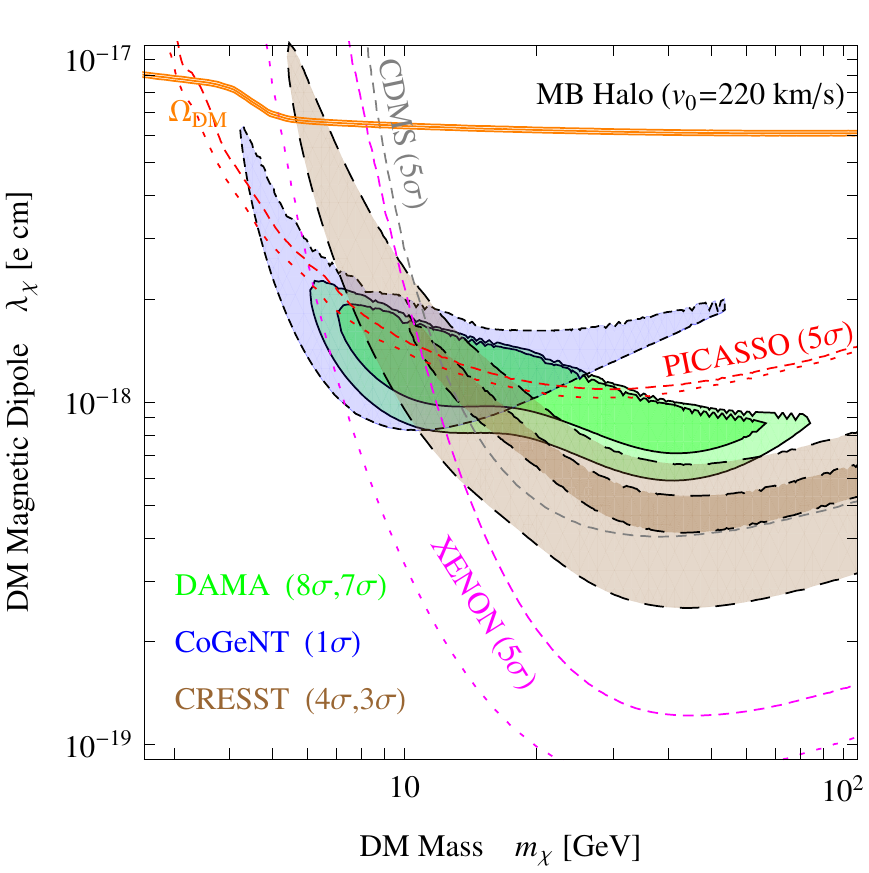}
\caption{DM magnetic dipole moment $\lambda_\chi$ as a function
of the Dark Matter mass $m_\chi$. The galactic halo has been assumed in the form of an isothermal
sphere with velocity dispersion $v_0=220$ km/s and local density
$\rho_0 = 0.3$ GeV/cm$^3$. Notations are the same as in Fig.~\ref{fig:CSI}; to match the two figures, one has to note that the role of $\sigma_p$ is played here by $\alpha \lambda_\chi^2$. 
The orange strip shows the values for $(m_\chi, \lambda_\chi)$ that fit the relic abundance $\Omega_{\mbox{\tiny DM}}$ assuming a completely thermal DM production (see Sec.~\ref{sec:omega}).}
\label{fig:B}
\end{figure}

\begin{figure}[!t]
\includegraphics[width=0.49\textwidth]{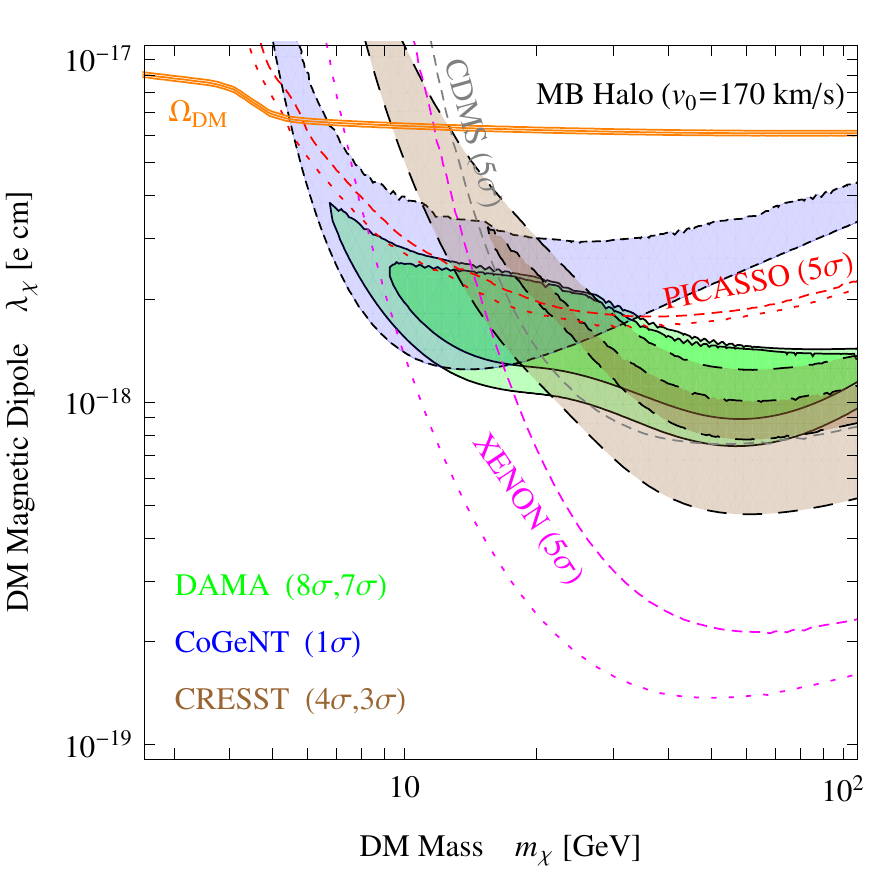}
\includegraphics[width=0.49\textwidth]{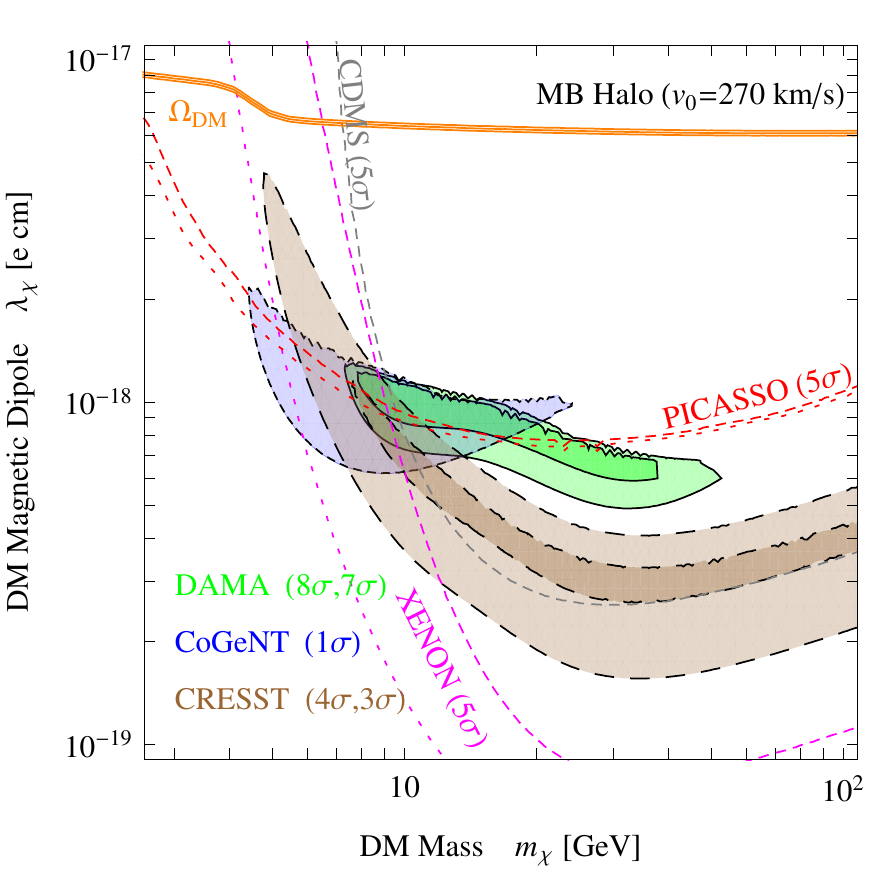}
\caption{DM magnetic dipole moment $\lambda_\chi$ as a function
of the Dark Matter mass $m_\chi$. The galactic halo has been assumed in the form of an isothermal
sphere with velocity dispersion $v_0=170$ km/s and local density
$\rho_0 = 0.18$ GeV/cm$^3$ (left panel); 
$v_0=270$ km/s and local density
$\rho_0 = 0.45$ GeV/cm$^3$ (right panel). Notations are the same as in Fig.~\ref{fig:CSI}; to match the two figures, one has to note that the role of $\sigma_p$ is played here by $\alpha \lambda_\chi^2$.%the matching in the two figures is provided by $\sigma_p \sim \alpha \lambda_\chi^2$
The orange strip shows the values for $(m_\chi, \lambda_\chi)$ that fit the relic abundance $\Omega_{\mbox{\tiny DM}}$, in the assumption of thermal DM production (see Sec.~\ref{sec:omega}).}
\label{fig:B-more}
\end{figure}

We analyze the direct detection data sets by using a standard isothermal halo model, which basically implies a truncated Maxwell-Boltzmann velocity distribution function (see discussion in Sec.~\ref{sec:nucrecrat}). Since the response of direct detection experiments is quite sensitive to the DM distribution in the galactic halo \cite{Belli:2002yt}, especially in the scenario we are interested to study, we take into account uncertainties
on the velocity dispersion $v_0$, as discussed in Ref.~\cite{Belli:2002yt}. 
%We recall that in the case of an isothermal model, the DM velocity dispersion is directly linked to the
%asymptotic value of the rotational velocity supported by the DM halo: uncertainties in the
%velocity dispersion $v_0$ are therefore representative of the uncertainties in the
%local rotational velocity \cite{Belli:2002yt}. 
We will use the three values $v_0 = 170, 220, 270$ km/s, which bracket the
uncertainty in the local rotational velocity. Let us notice that the value of the local DM density $\rho_0$ is correlated to the adopted value of $v_0$, as discussed e.g.~in Ref.~\cite{Belli:2002yt}. This corresponds to the model denoted
as A0 in \cite{Belli:2002yt}, and we adopt here the case of minimal halo, which implies
lower values of the local DM density (since a fraction of the galactic potential is supported
by the disk/bulge). In turn, this implies the adoption of $\rho_0 = 0.18, 0.30, 0.45$ GeV/cm$^3$ for $v_0 = 170, 220, 270$ km/s, respectively \cite{Belli:2002yt}.

Fig.~\ref{fig:B} shows the constraints and the favored regions coming from DM direct detection experiments in the $(m_\chi, \lambda_\chi)$ plane. The galactic halo has been assumed to be in the form of an isothermal
sphere with velocity dispersion $v_0=220$ km/s and local density
$\rho_0 = 0.3$ GeV/cm$^3$. %In this figure we show the allowed regions compatible
%with the annual modulation signal in DAMA, and the CoGeNT excess as well as the CRESST one, when interpreted as a DM signal. Specifically, 
The solid green contours denote the regions compatible with the DAMA annual modulation effect \cite{Bernabei:2008yi,Bernabei:2010mq}, in 
absence of channeling \cite{Bernabei:2007hw}. %The solid red contours B, refer to the regions compatible with the DAMA annual modulation effect, when the channeling effect is considered at its maximal value.\footnote{DAMA is a multi-target detector and allowed regions at large DM mass correspond to scattering on I, while at small DM mass regions correspond to scattering on Na. Note that the region at DM mass around 10 GeV is given by scattering on Na targets in the no-channeling case, and by scattering on I targets in the maximal channeling case.}
The short-dashed blue contour refers to the region derived from the CoGeNT annual modulation signal \cite{Aalseth:2011wp}, when the bound from the unmodulated CoGeNT data is taken into account. The dashed brown contours denote the regions compatible with the CRESST excess \cite{Angloher:2011uu}. 
For all the data sets, the contours refer to regions where the absence of modulation
can be excluded with a C.L. of $7\sigma$ (outer region), $8\sigma$ (inner region) for DAMA, 
$1\sigma$ for CoGeNT, and the absence of an excess can be excluded at $3\sigma$ (outer region), $4\sigma$ (inner region)
for CRESST. %The right panel, instead, shows a further analyses for the DAMA data: the solid orange contours refer to the results obtained by varying the channeling fraction $f_{\rm ch}$ in its allowed range, as discussed in Section \ref{sec:DAMA}. We can therefore see the extent of the DAMA allowed region
%when $f_{\rm ch}$ is marginalized over. We will adopt this procedure of treating
%the channeling effect in DAMA
%in the remainder of the paper. 
Constraints derived by the null result experiments are shown at $5\sigma$ as gray, magenta and red dashed lines for CDMS, XENON100 and PICASSO respectively. 
%The different lines refer to the various
%galactic halo models discusses in our analyses: broken lines refer to the isothermal 
%sphere with
%$v_0 = 170$ km/s and $\rho_0 = 0.18$ GeV/cm$^3$ (short-dashed line),
%$v_0 = 220$ km/s $\rho_0 = 0.3$ GeV/cm$^3$ (medium-dashed line),
%$v_0 = 270$ km/s and $\rho_0 = 0.45$ GeV/cm$^3$ (long-dashed line). 
%Solid lines refer to the triaxial halo
%model \cite{Evans:2000gr,Belli:2002yt}. 
For the XENON detector, as discussed in the previous section, the constraints
refer to thresholds of 4 and 8 photoelectrons, while for PICASSO we take into account the uncertainty in the intrinsic energy resolution of the employed detection technique \cite{Archambault:2012pm}. 
%In the first panel, the blue dashed line
%stands, instead, for a threshold of 4 photoelectron and for an isothermal sphere with
%$v_0 = 220$ km/s and $\rho_0 = 0.3$ GeV/cm$^3$. We can notice the extent
%of variation of the constraints when the galactic halo model and/or the mechanism
%of interaction is varied. 
%The value of 8 photoelectrons has been chosen in order to determine a situation which is nearly independent on the knowledge of $\mathcal{L}_{\rm eff}$ below 3 keVnr (for further details see \cite{FPR}). 
%Notice that other discussion and additional considerations on the assumption that one can be done to determine the XENON100 response to light DM can be found, e.g.~Refs.~\cite{Bernabei:2008jm,Collar:2011wq}, and thus due to the large uncertainties inherent in the derivation of bounds from XENON100 for light DM, it appears to be still preliminary to assume those bounds as strictly firm. For this reason, we consider the 8 PE bound as more appropriate since it is less dependent on the $\mathcal{L}_{\rm eff}$ extrapolation. 

We can see that, as expected from the discussion in Sec.~\ref{sec:theopred}, for a given value of $m_\chi$, all the experiments determine a DM-proton cross section $\sigma_p = \alpha \lambda_\chi^2$ that is higher than the one for the standard case of non isospin violating contact interaction depicted in Fig.~\ref{fig:CSI}; to give a benchmark value, DAMA points now to $\sigma_p %= \alpha \lambda_\chi^2 
\sim 1.5 \times 10^{-38}$ cm$^2$, about $10^2$ times above the standard case% \xxx{shown in Fig.~\ref{fig:CSI}}
, as shown in Sec.~\ref{sec:theopred}. Moreover we observe the expected gathering of the various experiments, that get closer to each other with respect to the standard scenario, for an overall better agreement, and with DAMA, CoGeNT and CRESST featuring a larger overlap at low masses. Both the DAMA and CoGeNT regions point towards a DM mass in the 10 GeV ballpark (more specifically, from about 7 up to about 12 GeV)
and DM magnetic dipole moment around $1.5 \times 10^{-18}$ $e \, \text{cm}$ without exceeding the constraints, corresponding to an inverse mass energy scale of circa $\Lambda_\text{M} \sim 10$ TeV. %\xxx{Interestingly, this value of $\lambda_\chi$ corresponds to an inverse mass energy scale of circa $\Lambda_\text{M} \sim 3$ TeV, surprisingly close to the electroweak natural energy scale of $4\pi \times 246$ GeV.}
CRESST allows for even heavier DM masses, but still compatible with the range determined by the other two experiments. CDMS and XENON, on the other hand, exclude a smaller part of the parameter space with respect to the standard case. These two experiments do not exclude the overlapping region once one accepts our conservative choice of the XENON threshold at 8 photoelectrons. At  $5\sigma$  it is seen that PICASSO cannot exclude the common regions of the experiments reporting a signal. 
Finally, we also notice the occurrence of the expected flattening of the various experiments, meaning that the tilt featured in the standard case (higher mass regions pointing to lower $\sigma_p$) is very suppressed in the case of magnetic moment interaction, as explained in Sec.~\ref{sec:theopred}.

These results depend on the galactic halo model assumed. The effect induced by the variation in the DM dispersion velocity is shown in the
two panels of Fig.~\ref{fig:B-more}. In the case of $v_0=170$ km/s, the regions are
not significantly modified as compared to the case $v_0=220$ km/s, except for the
overall normalization due to the different values of the local DM density
in the two cases. 
Larger dispersion velocities, instead, lead to the second regime discussed in Sec.~\ref{sec:theopred} also at small DM masses (see right panel of Fig.~\ref{fig:A}). Due to the high energy dependence of the differential rate in this regime, experiments that use heavy targets and/or have high thresholds, get a suppression of the event rate, and point therefore to relatively higher values of the interaction cross section compared to the standard case. Since the DM signal is expected at lower recoil energies, smaller DM masses are now favorite. The right panel of Fig.~\ref{fig:B-more} shows in fact that the allowed regions
are shifted toward smaller DM masses for $v_0=270$ km/s.

\section{Relic Abundance}
\label{sec:omega}

In addition to a satisfactory fitting of the direct detection data, the DM magnetic dipole interaction can accommodate the annihilation cross section appropriate for thermal production. Since it couples to the photon, the DM annihilates to any charged Standard Model particle-antiparticle pair, with mass lower than $m_\chi$. Considering a DM particle with 10 GeV mass, that as we argued fits well the direct detection experiments, {the allowed primary channels} are $e^+ e^-$, $\mu^+ \mu^-$, $\tau^+ \tau^-$, $u \bar u$, $d \bar d$, $s \bar s$, $c \bar c$, $b \bar b$ and $\gamma \gamma$.
\\
The DM annihilation cross sections to Standard Model fermions and photons at tree level are
\begin{multline}
\label{eq:fermionXsec}
\sigma_{\chi \bar\chi \rightarrow f \bar f}(s) =
\frac{ \alpha \, Q^2_f N_C^f }{12} \frac{\beta_f}{\beta_{\chi}} \frac{1}{s^2} \times
\\
\left[ \lambda^2_{\chi} \left( s^2 (3-\beta^2) +12 m^2_{\chi} s + 48 m^2_f m^2_{\chi} \right) + d^2_{\chi} \left( s^2 (3 -\beta^2) -12 m^2_{\chi} s \right) \right] \ ,
\end{multline}
\beq
\label{gammagamma}
\sigma_{\chi \bar\chi \rightarrow \gamma\gamma}(s) = \frac{\lambda_\chi^4}{4 \pi \beta_\chi} \left( m_\chi^2 + \frac{s}{8} - \frac{s \beta_\chi^2}{24} - 4 \frac{m_\chi^4}{s} \frac{\text{arcth}(\beta_\chi)}{\beta_\chi} \right) \ ,
%\frac{\lambda_\chi^4}{2 \pi \beta_\chi} \left( \frac{7}{12} m_\chi^2 + \frac{s}{24} - 2 \frac{m_\chi^4}{s} \frac{\text{arcth}(\beta_\chi)}{\beta_\chi} \right) \ ,
\eeq
where $\beta_{\chi, f} = (1 - 4 m_{\chi, f}^2/s)^{1/2}$, $\beta \equiv \beta_{\chi} \beta_f$,
$Q_f$ is the charge of the fermion and finally $N_C^\ell = 1$ for leptons and $N_C^q = 3$ for quarks. 
In the non-relativistic limit, Eqs.~\eqref{eq:fermionXsec} and \eqref{gammagamma} reduce to
\begin{align}
\langle \sigma_{\chi \bar\chi \rightarrow f \bar f} \ v_\text{rel} \rangle &\simeq N \, \alpha \lambda^2_{\chi} \cdot \text{BR}(f) \ ,
%\beq
%\sigma_\text{ann} v_\text{rel} \simeq N \alpha \lambda^2_{\chi} \ ,
%\eeq
\\
\label{gammagammaNR}
\langle \sigma_{\chi \bar\chi \rightarrow \gamma\gamma} \ v_\text{rel} \rangle &\simeq \frac{1}{4 \pi} \lambda_\chi^4 m_\chi^2 \ ,
\end{align}
where $N = \sum_f Q^2_f N_C^f = 20/3$ accounts for the number of degrees of freedom of the Standard Model fermions into which the DM can annihilate; the branching ratios $\text{BR}(f)$ are defined as $\text{BR}(f) = Q^2_f N_C^f / N$. {As one can see, for the low values of $\lambda_\chi m_\chi$ pointed by the direct detection experiments, the two photons final state is suppressed with respect to the fermionic one.} Using a DM energy density $\Omega_{\mbox{\tiny DM}} h^2 =0.1126 \pm 0.0036$ measured by WMAP~\cite{Komatsu:2010fb}, we show\footnote{The figures are produced by solving numerically the Boltzmann equation

\begin{equation}
\dot n_{\rm tot}+3H n_{\rm tot}=-\frac12 \langle \sigma_\text{ann} v_\text{rel} \rangle\left(n_{\rm tot}^2-n_{\rm eq}^2\right),
\end{equation}
where $ n_{\rm tot}$ is the total number density of particles and antiparticles, $H$ is the Hubble parameter and for $\langle \sigma_\text{ann} v_\text{rel} \rangle$ we use the more precise formula given in Eq.~(3.8) of \cite{Gondolo:1990dk}.}
%\begin{equation}
% \langle \sigma_\text{ann} v_\text{rel} \rangle = 
 %      \frac{1}{8m_{\chi}^{4}TK^{2}_2(\frac{m_{\chi}}{T})}
 %     \int_{4m_\chi^2}^{\infty} ds
 %                      \sqrt{s}(s-4m_{\chi}^2)K_1(\frac{\sqrt{s}}{T})
 %                      \sigma_{\rm tot}(s) \ ,
%\end{equation}
%where $K_1$ and $K_2$ are modified Bessel functions of the second kind and $s$ is the Mandelstam invariant; for the total cross section $\sigma_{\rm tot}(s)$ we considered $\sum_f \sigma_{\chi \bar\chi \rightarrow f \bar f}(s)$
with an orange strip in the $(m_\chi, \lambda_\chi)$ plane of Figs.~\ref{fig:B} and \ref{fig:B-more} the phase space where a thermal production of the magnetic DM is possible.
 %\footnote{Here $\Omega_{\mbox{\tiny DM}} = \rho_{\mbox{\tiny DM}}/\rho_c$ is defined as usual as the DM energy density divided by the critical energy density of the Universe $\rho_c = 3 H_0^2/8\pi G_N$, where $H_0$ is the present Hubble parameter. $h$ is its reduced value \mbox{$h = H_0 / 100$ km/s Mpc$^{-1}$}.}. 
%Our result\footnote{The figures are produced by solving numerically the more precise formula~\cite{Gondolo:1990dk}
%\begin{equation}
%   \langle \sigma_\text{ann} v\rangle = 
   %    \frac{1}{8m_{\chi}^{4}TK^{2}_2(\frac{m_{\chi}}{T})}
     % \int_{4m_2^2}^{\infty} ds
         %               \sqrt{s}(s-4m_{\chi}^2)K_1(\frac{\sqrt{s}}{T})
            %            \sigma_{\rm tot}(s) \ ,
%\end{equation}
%where $K_1$ and $K_2$ are modified Bessel functions of the second kind and $s$ is the Mandelstam invariant; for the total cross section $\sigma_{\rm tot}(s)$ we considered $\sum_f \sigma_{\chi \chi \rightarrow f \bar f}(s)$.} is shown as an orange strip in Figs.~\ref{fig:B} and \ref{fig:B-more}; we checked that what we find is in agreement with the result of Ref.~\cite{Sigurdson:2004zp}.

It can be seen in Figs.~\ref{fig:B} and \ref{fig:B-more} that the agreement between the allowed regions and the relic abundance depends strongly on the dispersion velocity $v_0$. The best agreement takes place for lower values of $v_0$. This is due to the fact that, as commented already in Sec.~\ref{sec:results}, $\rho_0$ is linked to the dispersion velocity $v_0$. %such that $\rho_0 / v_0^2$ is fixed. 
By varying $v_0$, $\rho_0$ is modified accordingly, changing therefore the expected rate and the fitting values for the interaction cross sections, or in other words $\lambda_\chi$. As a consequence, the favored regions and constraint lines of direct detection experiments move along the vertical axis in the $(m_\chi, \lambda_\chi)$ plane. For lower velocity dispersion, both the local DM density and  the expected rate decrease. Therefore, the favored regions will point to a higher DM magnetic moment, getting closer to the relic abundance strip. The value of the relic abundance is obviously independent on the local DM density.

\bigskip

The fact that the magnetic DM (with only two parameters) provides a good fit to the direct detection experiments and simultaneously can accommodate a thermal annihilation cross section (which is several orders of magnitude stronger than the DM-nucleon cross section in the context of ``standard'' contact interaction) is not easily met by other DM candidates. Although there are  candidates that can have a DM-nucleon cross section much lower than their thermal annihilation cross section, these candidates have usually spin-dependent interactions and cannot fit nicely the direct search experiments with a positive DM signal.
%feature of providing a good fit to the relic abundance, at the same time with fitting the direct detection experiments, is a natural property of the magnetic dipole DM not \xxx{shared with many/easy to obtain in} other DM models.
For a typical candidate with spin-independent contact interactions, and for a strength of interaction that leads to a DM-nucleon cross section pointed by the direct detection experiments featuring a signal, the annihilation cross section is way too small to produce this candidate thermally. In order to match, a suppression mechanism for the direct detection event rate should take place. In the case of magnetic moment DM, this mechanism exists and has two different reasons for the SI and the SD parts of the interaction. As already explained in Sec.~\ref{sec:theopred}, this suppression is encoded into the function $\Theta$ in Eq.~\eqref{Theta}, that spans roughly from $10^{-3}$ to $10^{-1}$ for a 10 GeV DM (see Fig.~\ref{fig:A}). This suppression is a result between two competing factors for the SI case: the enhancement provided by the $E_R^{-1}$ dependence of the differential cross section, and the suppression provided by the kinetic integral; in the SD case, the suppression is instead due to the lack of the $A^2$ enhancement usually present in the standard SI case.

Notice that in the case of electric dipole DM one faces a very different situation: here the differential cross section has the same dependence on the velocity as in a contact interaction (Eq.~\eqref{diffCSE}), and therefore there is no kinetic suppression, while on the contrary the rate is enhanced by the $E_R^{-1}$ dependence. Moreover, the annihilation to fermions is a p-wave process, Eq.~\eqref{eq:fermionXsec}, and therefore the annihilation cross section is suppressed, making the value of $d_\chi$ needed to fit the relic abundance even bigger.

Furthermore, it is worth to point out that neither asymmetric/mixed DM \cite{Belyaev:2010kp}, nor oscillating DM \cite{Cirelli:2011ac,Tulin:2012re,Buckley:2011ye} can improve the agreement between the relic density and the allowed regions of direct DM searches in the context of magnetic DM interaction.

%There is also the intriguing possibility that magnetic moment DM is mixed, i.e. has a symmetric and an asymmetric component  like the one investigated in \cite{Belyaev:2010kp}.   For the mixed and fully asymmetric DM one will need, however, to further deplete  the symmetric component by adding new interactions. 

\section{Constraints}
\label{sec:constraints}

Given the qualitative agreement between the direct detection allowed regions and the fit of the relic abundance, it is natural to ask whether constraints coming from other searches could limit the parameter space of the magnetic moment DM. This is even more pressing since the bounds coming from indirect DM searches are on the verge of constraining the thermal annihilation rate for light DM particles.
We will discuss the constraints  imposed by indirect searches, colliders and by the observations of compact stars,  and then identify the most stringent ones.

%coming from the different epochs,

\begin{figure}[!t]
\centering
\includegraphics[width=.6\textwidth]{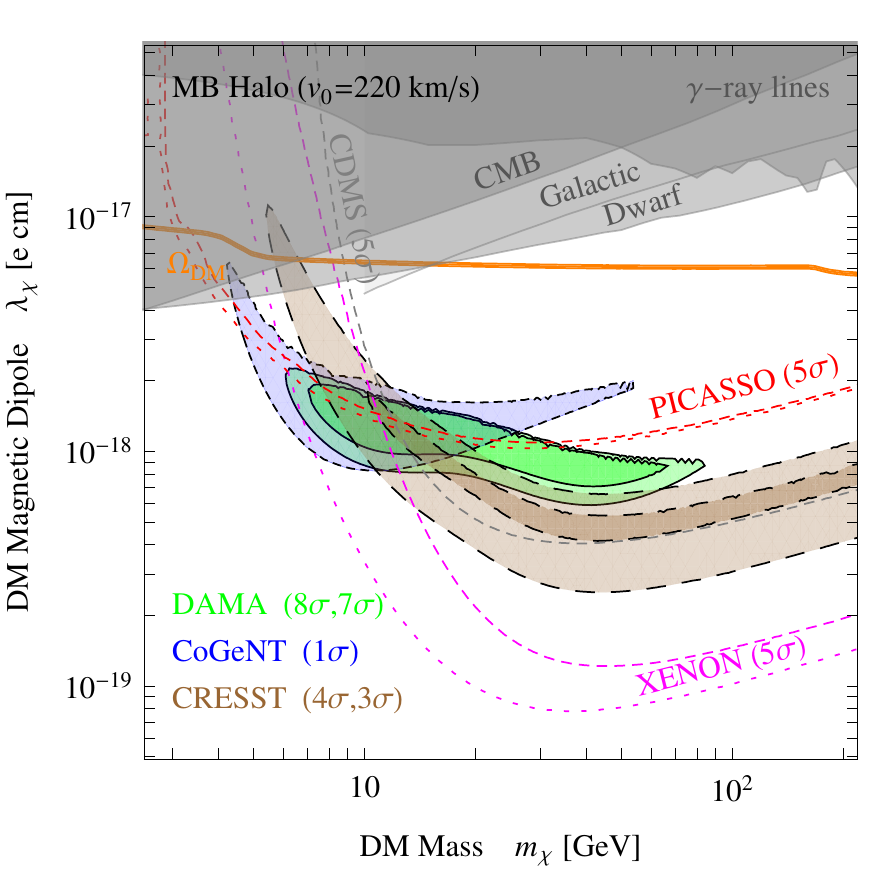}
\caption{DM magnetic dipole moment $\lambda_\chi$ as a function
of the Dark Matter mass $m_\chi$. The galactic halo has been assumed in the form of an isothermal
sphere with velocity dispersion $v_0=220$ km/s and local density
$\rho_0 = 0.3$ GeV/cm$^3$. Notations are the same as in Fig.~\ref{fig:CSI}%; to match the two figures, one has to note that the role of $\sigma_p$ is played here by $\alpha \lambda_\chi^2$%the matching in the two figures is provided by $\sigma_p \sim \alpha \lambda_\chi^2$
. The orange strip shows the values for $(m_\chi, \lambda_\chi)$ that fit the relic abundance $\Omega_{\mbox{\tiny DM}}$, in the assumption of thermal DM production (see Sec.~\ref{sec:omega}). The shaded regions refer to the constraints from $\gamma$-ray lines, CMB, galactic $\gamma$-rays and dwarf galaxies. These constraints are only valid in the assumption of symmetric DM. The galactic photons constraint enforces total annihilation of the DM into $b \bar b$, and therefore a less stringent constraint is expected for magnetic moment DM.}
\label{fig:constraints}
\end{figure}

\subsection{Epoch of reionization and CMB}

Strong constraints are imposed on DM annihilations from considering the effect on the generation of the CMB anisotropies at the epoch of recombination (at redshift $\sim1100$) and their subsequent evolution down to the epoch of reionization. The actual physical effect of energy injection around the recombination epoch results in an increased amount of free electrons, which survive to lower redshifts and affect the CMB anisotropies~\cite{Galli:2009zc,Slatyer:2009yq,Huetsi:2009ex,Cirelli:2009bb}. Detailed constraints have been recently derived in~\cite{Hutsi:2011vx,Galli:2011rz}, based on the WMAP (7-year) and Atacama Cosmology Telescope 2008 data. The constraints are somewhat sensitive to the dominant DM annihilation channel: annihilation modes for which a portion of the energy is carried away by neutrinos or stored in protons have a smaller impact on the CMB; on the contrary the annihilation mode which produces directly $e^+e^-$ is the most effective one. Usually the approach here is to consider $100\%$ annihilation rate in a single final state; anyway, in our case several annihilation channels are open, and therefore we expect a smaller energy injection in the interstellar medium with respect to the case of annihilation only into $e^+e^-$. In Fig.~\ref{fig:constraints} we reproduce the constraints as obtained in Refs.~\cite{Cirelli:2009bb,Hutsi:2011vx,Galli:2011rz} considering now all the channels, each with its branching ratio defined in Sec.~\ref{sec:omega}.% As expected, our constraint for a $10$ GeV DM particle does not exclude the value of $\lambda_\chi$ needed to explain a thermal production.

\subsection{Present epoch $\mathbf{\gamma}$-rays}

For most of the DM annihilation modes, another relevant constraint is in fact imposed by the indirect DM searches in the present epoch. The DM constraints provided by the FERMI-LAT $\gamma$-ray data are particularly relevant as they are now cutting into the thermal annihilation cross section for low DM masses ($\lesssim 30$ GeV) and a variety of channels.

In particular, dwarf satellite galaxies of the Milky Way are among the most promising targets for Dark Matter searches in $\gamma$-rays because of their large dynamical mass to light ratio and small expected background from astrophysical sources. No dwarf galaxy has been detected in $\gamma$-rays so far and stringent upper limits are placed on DM annihilation by applying a joint likelihood analysis to 10 satellite galaxies with 2 years of FERMI-LAT data, taking into account the uncertainty in the Dark Matter distribution in the satellites~\cite{Ackermann:2011wa}. The limits are particularly strong for hadronic annihilation channels, and somewhat weaker for leptonic channels as diffusion of leptons out of these systems is poorly constrained. In our case, having both hadronic as well as leptonic annihilation channels, we expect again a weaker constraint with respect to the pure hadronic annihilation (see e.g.~figure 2 in Ref.~\cite{Ackermann:2011wa}).

Other strong limits on annihilation channels are set by, for example, the $\gamma$-ray diffuse emission measurement at intermediate latitudes, which probes DM annihilation in our Milky Way halo \cite{Cirelli:2009dv,Papucci:2009gd,Zaharijas:2010ca}. In particular, the most recent limits come from 2 years of the FERMI-LAT data in the $5^{\circ} \leqslant b \leqslant 15 ^{\circ}$, $-80^{\circ} \leqslant \ell \leqslant 80^{\circ}$ region \cite{Zaharijas:2010ca}, where $b$ and $\ell$ are the galactic latitude and longitude. Since bounds on all annihilation channels are not available with the latest data, we report only the most stringent one, coming from $b \bar b$.

The last kind of constraints that we can set on DM annihilation in the case of magnetic moment interaction comes from $\gamma$-ray lines. Indeed, given the possibility of annihilation in two photons, we expect a line in the cosmic photon spectrum at energies equal to the mass of the DM. Constraints on the annihilation cross section into photons can be drawn from the latest FERMI-LAT data \cite{Abdo:2010nc}. For instance, the FERMI-LAT collaboration excludes at $95\%$ the annihilation $\langle \sigma_{\chi \bar\chi \rightarrow f \bar f} \ v_\text{rel} \rangle < 0.5 \times 10^{-27}$ cm$^3$/s for a DM mass $m_\chi = 30$ GeV in the case of an isothermal halo. In our case we use the latest (preliminary) data on the flux from spectral lines (shown in slide 39 of \cite{Morselli}) to constrain the annihilation cross section of Eq.~\eqref{gammagammaNR}\footnote{A similar study has been published in Ref.~\cite{Goodman:2010qn} for DM masses above $30$ GeV.}. We consider  an all-sky region with the Galactic plane removed ($|b|>10^\circ$), plus a $20^\circ \times 20^\circ$ square region centered on the Galactic center \cite{Abdo:2010nc}. Notice that, for small DM masses and dipole moments, the one-loop contribution becomes comparable with the tree level one; a rough estimate of the loop cross section in the non-relativistic limit is $\alpha^3 \lambda_\chi^2 / 4 \pi$, and therefore we expect it to become sizeable for $\lambda_\chi m_\chi < \alpha^{3/2} \simeq 6 \times 10^{-4}$.  
In the part of the parameter space probed by $\gamma$-ray lines searches the loop correction is negligible.

\bigskip

We superimpose all these constraints in Fig.~\ref{fig:constraints}. We see that, apart from the $\gamma$ lines bound, they are somewhat stronger than the CMB one considered above. We keep however the latter as it is less model dependent.% More generally, we stress that all these constraints are valid under different assumptions, such as e.g.~the DM galactic profile. As a consequence, while we report all of them on the same plane for completeness, the precise regions of the parameter space which are actually excluded depend on the precise DM model.

As shown in Fig.~\ref{fig:B-more}, a change in the DM local density modifies the direct searches results. This does not happen for indirect searches, as they are not very sensitive to $\rho_0$. Therefore the constraints shown in Fig.~\ref{fig:constraints} also apply, unchanged, to this case. For lower DM dispersion velocity $v_0$ these constraints start to play an important role in cutting the parameter space for direct detection searches.
%A change the DM local density modifies the direct searches results as shown in Fig.~\ref{fig:B-more}, but we expect the constraints coming from indirect detection experiments not to change much. The CMB and dwarf galaxies constraints do not depend of course on the galactic DM profile, while the bounds from diffuse gamma-rays and gamma lines outside the galactic center depend on an average val

\subsection{Collider and other Astrophysical constraints}

The constraints we have addressed above apply to the case of symmetric (thermally produced) DM. These constraints are not relevant in the case where DM is of asymmetric nature. There are two extra type of constraints: constraints imposed by collider searches, and constraints imposed by observations of compact stars such as white dwarfs and neutron stars. The collider constraints are applicable wether DM is symmetric or not, whereas the compact star constraints are valid only for asymmetric DM.  

The collider constraints emerge from the fact that for a given  $\lambda_{\chi}$ and $m_{\chi}$ that fit the direct DM searches experimental data (and even provide the proper DM annihilation cross section for thermal production), the cross section for production of a pair of DM particles in colliders is completely fixed. The processes that lead to constraints are mono-photon production (from initial or final state) plus missing energy due to the pair of DM particles in $e^+e^-$ collisions at LEP, or mono-jet production plus missing energy in proton-antiproton collision in Tevatron, or proton-proton collisions at LHC. In the case of magnetic DM these constraints have been studied in~\cite{Fortin:2011hv, Barger:2012pf} where it is found that the upper bound on $\lambda_{\chi}$ is safely above the range of values of $\lambda_{\chi}$ relevant for the direct search experiments (for the range of $m_{\chi}\sim 10$ GeV).

In the case of asymmetric DM, constraints can be imposed by compact star observations based on the fact that a substantial number of captured DM  particles might lead to gravitational collapse and formation of a black hole that can destroy the host star. The magnetic DM being a fermion evades the severe constraints on asymmetric bosonic DM based on neutron stars \cite{Kouvaris:2011fi,McDermott:2011jp,Guver:2012ba}. The constraints on self-DM cross section with Yukawa interactions presented in \cite{Kouvaris:2011gb} are avoided for several reasons: firstly DM-DM interactions scale as $m_{\chi}^2 \lambda_{\chi}^4$ (leading to a typical DM-DM cross section of $\sim10^{-43}$ cm$^2$ which is much smaller than the constraint). Secondly the constraints are not directly applicable because the mediator of the magnetic DM is a massless photon (and not a massive mediator necessary for the constraint) and the DM-DM interaction is repulsive. The latter adds up to the effect of the Fermi pressure of the DM particles and therefore the amount of particles needed for gravitational collapse cannot be accumulated within the age of the Universe. A potential attractive photon interaction takes place in the symmetric case, which will however lead also to annihilation of the DM population inside the neutron star invalidating thus the constraint derived from black hole formation. 

Finally the magnetic DM evades constraints on the spin-dependent part of the cross section imposed by observations of white dwarfs~\cite{Kouvaris:2010jy}. Although these constraints are typically weaker than the ones derived from direct searches at the range $m_{\chi}\sim10$ GeV, they could become stricter if spin-dependent interactions scale as some (positive) power of the recoil energy. Since DM particles acquire high velocities when entering the white dwarf, such constraints could in principle exclude such a candidate. However, as it was pointed out in~\cite{Barger:2010gv}, the spin-dependent cross section does not scale with the recoil energy and therefore the white dwarf constraints can be safely ignored. 

\section{Conclusions}
\label{sec:conclusions}

We investigated a fermionic Dark Matter particle carrying magnetic dipole moment and analyzed its impact on direct detection experiments. We provided an analytic understanding of how the photon induced long-range interaction neatly modifies the experimental allowed regions with respect to the typically assumed contact interaction. We showed that this candidate can accommodate the DAMA, CoGeNT and CRESST experimental results. By assuming conservative bounds we have demonstrated that this candidate is not ruled out by the CDMS, XENON and PICASSO experiments. Assuming a symmetric Dark Matter sector, we also determined the associated thermal relic density and provided relevant bounds from indirect search experiments. We found that all the experimental results are compatible with a Dark Matter particle with mass around $10$ GeV and magnetic moment $1.5 \times 10^{-18}$ $e \, \text{cm}$, which corresponds to a new physics scale $\Lambda_\text{M} \sim 10$ TeV.

\bigskip

\noindent {\bf Acknowledgments.}  We thank Robert Foot, Aldo Morselli and Kimmo Kainulainen for useful discussions. We are particularly grateful to Marco Regis for sharing parts of the code used for the data analysis.

\appendix
\section{Appendix}
\label{appendix}

In this appendix we estimate the effect of the dipole-dipole interaction term in Eq.~\eqref{diffCSM} for all the targets considered in our analysis. %In the following we explain in more detail why in the context of elastic Dark Matter (DM) the spin-independent (SI) interaction is negligible with respect to the spin-dependent (SD) one, for all the targets considered in our analysis. 
To this aim, one can define the useful ratio
\beq
\label{ratio}
r(E_R) \equiv \frac{dR_\text{SD}}{dE_R} \cdot \left(\frac{dR_\text{SI}}{dE_R} \right)^{-1}=\frac{\int d\sigma_\text{SD}/{dE_R} \,v\, dn_\chi}{\int d\sigma_\text{SI}/{dE_R} \,v \,dn_\chi} \ ,
\eeq
which measures the main contribution of the total recoil rate for a given nuclear specie in a detector. If $r>1$ the interaction is largely spin-dependent, while it is mostly spin-independent for $r< 1$.

\begin{figure*}[h!]
\centering
\includegraphics[width=.49\textwidth]{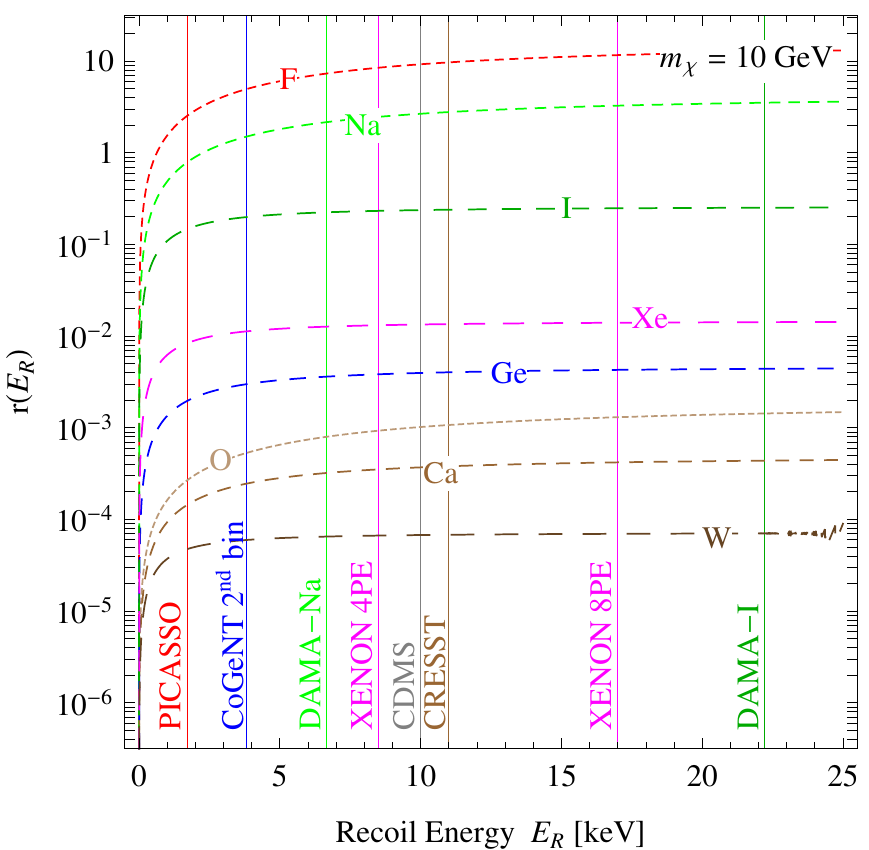}
\includegraphics[width=.49\textwidth]{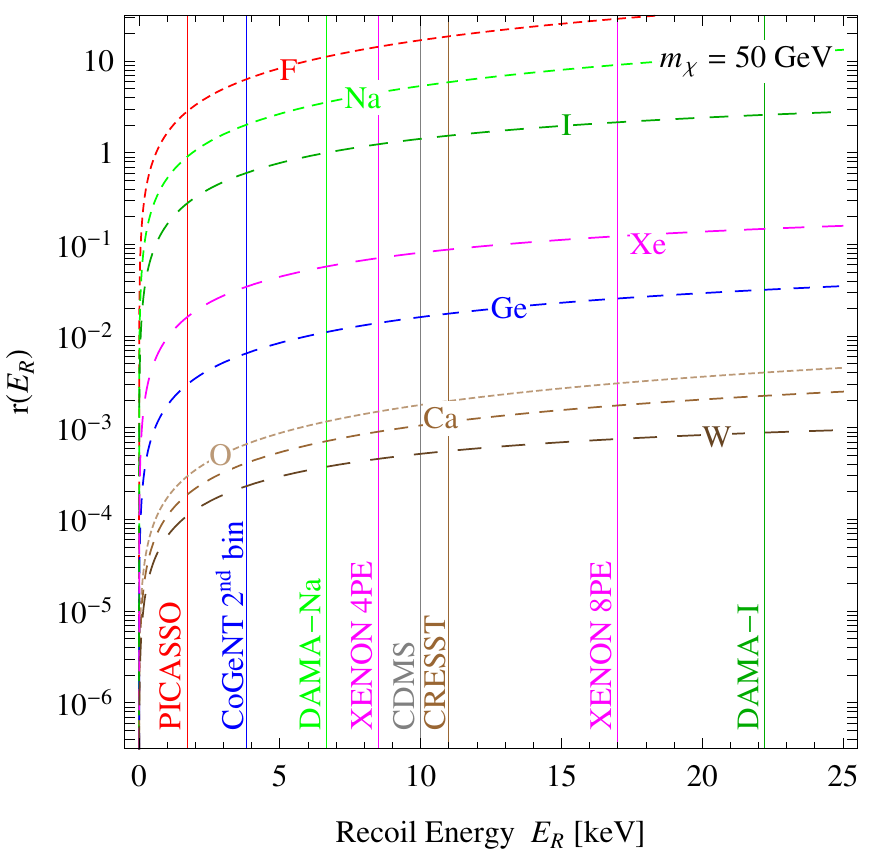}
\caption{The SD to SI rates ratio $r$ defined in Eq.~\eqref{ratio} for the two cases of $m_\chi = 10$ GeV (left) and $m_\chi = 50$ GeV (right). We consider here a Maxwellian halo with local dispersion velocity $220$ km/s. Supposing the major part of the signal to come from the lower energy threshold (or from the second bin in the case of CoGeNT), $r$ can be determined in the figure by the point where the dashed line meets the vertical line with the same color, for each nuclear element.}
\label{r-fig}
\end{figure*}

For elastic DM, which features $\delta=0$, the ratio $r$ can be computed from Eqs.~\eqref{ThetaSI} and~\eqref{ThetaSD}, yielding
\beq
r(E_R) = \frac 23 \, \Theta_{\rm SI}(E_R)^{-1} \left( \frac{1}{A} \frac{\bar \lambda_{\rm nuc}}{\lambda_p} \frac{\mu_{\chi n}}{m_p} \right)^2 \ ,
\eeq
where for the illustrative purposes of this appendix we approximated here $F_\text{SI}(E_R) \simeq F_\text{SD}(E_R)$. In Fig.~\ref{r-fig} we show the ratio $r$ as a function of the nuclear recoil energy considering a DM mass of $10$ GeV, which gives a good combined fit to the existing experiments, and $50$ GeV, for a comparison. % For each element we weighted over the stable isotopes. %The highest values of $r$ are obtained for the lightest elements, namely PICASSO ($r \simeq 0.3$) and DAMA-Na ($r \simeq 0.2$ for $m_\chi = 10$ GeV, $r \simeq 0.3$ for $m_\chi = 50$ GeV). Despite the high magnetic moment of Iodine nuclei, heavy elements feature an even smaller value of $r$; the fact that the SD DM-Iodine scattering cross section plays an important role in \hhref{1007.4200} is therefore to ascribe exclusively to the dynamics of the inelastic scattering. Even a $30\%$ error on the rate accounts anyway only for a $\sim 14\%$ error on $\mu_\chi$, since the rate is proportional to $\mu_\chi^2$. Moreover the positive SD contribution to the rate, even if not decisive for our result, helps with a better fit since it lowers slightly the DAMA region at low $m_\chi$. The PICASSO bound, on the other side, while becoming slightly more stringent, doesn't improve enough to pose a threat to the experiments featuring positive results.
One can see that the dipole-dipole term plays a major role for DAMA (Na, I) and PICASSO (F), while being negligible for all the other experiments considered, as stated in the main text.

While the dipole-charge interaction already provides a good fit, the effect of the dipole-dipole interaction contributes to a better agreement between the experiments. In fact, the increase in the cross section in DAMA favors a lower value for the DM magnetic moment $\lambda_\chi$, and therefore the DAMA region shifts down towards CoGeNT and CRESST. The overlap becomes then almost perfect, for a Maxwellian velocity distribution, especially with velocity dispersion $220$ km/s. On the other hand also the bound by PICASSO lowers, but even this enhanced constraint excludes only a minor part of the overlap zone.


\newpage


\begin{thebibliography}{nn}

\bibitem{Bernabei:1998fta}
  R.~Bernabei, P.~Belli, F.~Montecchia, W.~Di Nicolantonio, A.~Incicchitti, D.~Prosperi, C.~Bacci and C.~J.~Dai {\it et al.},
  %``Searching for WIMPs by the annual modulation signature,''
  Phys.\ Lett.\ B {\bf 424} (1998) 195.
  %%CITATION = PHLTA,B424,195;%%

\bibitem{Drukier:1986tm}
  A.~K.~Drukier, K.~Freese and D.~N.~Spergel,
  %``Detecting Cold Dark Matter Candidates,''
  Phys.\ Rev.\ D {\bf 33} (1986) 3495.
  %%CITATION = PHRVA,D33,3495;%%

\bibitem{Freese:1987wu}
  K.~Freese, J.~A.~Frieman and A.~Gould,
  %``Signal Modulation in Cold Dark Matter Detection,''
  Phys.\ Rev.\ D {\bf 37} (1988) 3388.
  %%CITATION = PHRVA,D37,3388;%%

\bibitem{Bernabei:2008yi}
  R.~Bernabei {\it et al.}  [DAMA Collaboration],
  %``First results from DAMA/LIBRA and the combined results with DAMA/NaI,''
  Eur.\ Phys.\ J.\ C {\bf 56} (2008) 333
  [arXiv:0804.2741 [astro-ph]].
  %%CITATION = ARXIV:0804.2741;%%

\bibitem{Bernabei:2010mq}
  R.~Bernabei, P.~Belli, F.~Cappella, R.~Cerulli, C.~J.~Dai, A.~d'Angelo, H.~L.~He and A.~Incicchitti {\it et al.},
  %``New results from DAMA/LIBRA,''
  Eur.\ Phys.\ J.\ C {\bf 67} (2010) 39
  [arXiv:1002.1028 [astro-ph.GA]].
  %%CITATION = ARXIV:1002.1028;%%

\bibitem{Belli:2002yt}
  P.~Belli, R.~Cerulli, N.~Fornengo and S.~Scopel,
  %``Effect of the galactic halo modeling on the DAMA / NaI annual modulation result: an Extended analysis of the data for WIMPs with a purely spin independent coupling,''
  Phys.\ Rev.\ D {\bf 66} (2002) 043503
  [hep-ph/0203242].
  %%CITATION = HEP-PH/0203242;%%

\bibitem{Belli:2011kw}
  P.~Belli, R.~Bernabei, A.~Bottino, F.~Cappella, R.~Cerulli, N.~Fornengo and S.~Scopel,
  %``Observations of annual modulation in direct detection of relic particles and light neutralinos,''
  Phys.\ Rev.\ D {\bf 84} (2011) 055014
  [arXiv:1106.4667 [hep-ph]].
  %%CITATION = ARXIV:1106.4667;%%

\bibitem{Aalseth:2010vx}
  C.~E.~Aalseth {\it et al.}  [CoGeNT Collaboration],
  %``Results from a Search for Light-Mass Dark Matter with a P-type Point Contact Germanium Detector,''
  Phys.\ Rev.\ Lett.\  {\bf 106} (2011) 131301
  [arXiv:1002.4703 [astro-ph.CO]].
  %%CITATION = ARXIV:1002.4703;%%

\bibitem{Aalseth:2011wp}
  C.~E.~Aalseth, P.~S.~Barbeau, J.~Colaresi, J.~I.~Collar, J.~Diaz Leon, J.~E.~Fast, and N.~Fields %and T.~W.~Hossbach 
  {\it et al.},
  %``Search for an Annual Modulation in a P-type Point Contact Germanium Dark Matter Detector,''
  Phys.\ Rev.\ Lett.\  {\bf 107} (2011) 141301
  [arXiv:1106.0650 [astro-ph.CO]].
  %%CITATION = ARXIV:1106.0650;%%

\bibitem{Angloher:2011uu}
  G.~Angloher, M.~Bauer, I.~Bavykina, A.~Bento, C.~Bucci, C.~Ciemniak, G.~Deuter and F.~von Feilitzsch {\it et al.},
  %``Results from 730 kg days of the CRESST-II Dark Matter Search,''
  arXiv:1109.0702 [astro-ph.CO].
  %%CITATION = ARXIV:1109.0702;%%

\bibitem{Foot:2011pi}
  R.~Foot,
  %``Mirror and hidden sector dark matter in the light of new CoGeNT data,''
  Phys.\ Lett.\ B {\bf 703} (2011) 7
  [arXiv:1106.2688 [hep-ph]].
  %%CITATION = ARXIV:1106.2688;%%

\bibitem{Schwetz:2011xm}
  T.~Schwetz, J.~Zupan,
  %``Dark Matter attempts for CoGeNT and DAMA,''
  JCAP {\bf 1108 } (2011)  008.
  [arXiv:1106.6241 [hep-ph]].
  
\bibitem{Farina:2011pw}
  M.~Farina, D.~Pappadopulo, A.~Strumia and T.~Volansky,
  %``Can CoGeNT and DAMA Modulations Be Due to Dark Matter?,''
  JCAP {\bf 1111} (2011) 010
  [arXiv:1107.0715 [hep-ph]].
  %%CITATION = ARXIV:1107.0715;%%

\bibitem{McCabe:2011sr}
  C.~McCabe,
  %``DAMA and CoGeNT without astrophysical uncertainties,''
  Phys.\ Rev.\ D {\bf 84} (2011) 043525
  [arXiv:1107.0741 [hep-ph]].
  %%CITATION = ARXIV:1107.0741;%%

\bibitem{Fox:2011px}
  P.~J.~Fox, J.~Kopp, M.~Lisanti, N.~Weiner,
  %``A CoGeNT Modulation Analysis,''
    [arXiv:1107.0717 [hep-ph]].

\bibitem{Hooper:2011hd}
  D.~Hooper and C.~Kelso,
  %``Implications of CoGeNT's New Results For Dark Matter,''
  Phys.\ Rev.\ D {\bf 84} (2011) 083001
  [arXiv:1106.1066 [hep-ph]].
  %%CITATION = ARXIV:1106.1066;%%

\bibitem{Gondolo:2011eq}
  P.~Gondolo, P.~Ko and Y.~Omura,
  %``Light dark matter in leptophobic Z' models,''
  Phys.\ Rev.\ D {\bf 85} (2012) 035022
  [arXiv:1106.0885 [hep-ph]].
  %%CITATION = ARXIV:1106.0885;%%

\bibitem{DelNobile:2011je}
  E.~Del Nobile, C.~Kouvaris, F.~Sannino,
  %``Interfering Composite Asymmetric Dark Matter for DAMA and CoGeNT,''
  Phys.\ Rev.\  {\bf D84 } (2011)  027301.
  [arXiv:1105.5431 [hep-ph]].
  
\bibitem{Arina:2011si}
  C.~Arina, J.~Hamann and Y.~Y.~Y.~Wong,
  %``A Bayesian view of the current status of dark matter direct searches,''
  JCAP {\bf 1109} (2011) 022
  [arXiv:1105.5121 [hep-ph]].
  %%CITATION = ARXIV:1105.5121;%%

\bibitem{Frandsen:2011ts}
  M.~T.~Frandsen, F.~Kahlhoefer, J.~March-Russell, C.~McCabe, M.~McCullough, K.~Schmidt-Hoberg,
  %``On the DAMA and CoGeNT Modulations,''
  [arXiv:1105.3734 [hep-ph]].
  
\bibitem{Kaplan:2011yj}
  D.~E.~Kaplan, G.~Z.~Krnjaic, K.~R.~Rehermann and C.~M.~Wells,
  %``Dark Atoms: Asymmetry and Direct Detection,''
  arXiv:1105.2073 [hep-ph].
  %%CITATION = ARXIV:1105.2073;%%

\bibitem{Feng:2011vu}
  J.~L.~Feng, J.~Kumar, D.~Marfatia and D.~Sanford,
  %``Isospin-Violating Dark Matter,''
  Phys.\ Lett.\ B {\bf 703} (2011) 124
  [arXiv:1102.4331 [hep-ph]].
  %%CITATION = ARXIV:1102.4331;%%

\bibitem{Fitzpatrick:2010br}
  A.~L.~Fitzpatrick and K.~M.~Zurek,
  %``Dark Moments and the DAMA-CoGeNT Puzzle,''
  Phys.\ Rev.\  D {\bf 82} (2010) 075004
  [arXiv:1007.5325 [hep-ph]].
  %%CITATION = PHRVA,D82,075004;%%

\bibitem{Hooper:2010uy}
  D.~Hooper, J.~I.~Collar, J.~Hall, D.~McKinsey and C.~Kelso,
  %``A Consistent Dark Matter Interpretation For CoGeNT and DAMA/LIBRA,''
  Phys.\ Rev.\  D {\bf 82} (2010) 123509
  [arXiv:1007.1005 [hep-ph]].
  %%CITATION = PHRVA,D82,123509;%%

\bibitem{Foot:2010rj}
  R.~Foot,
  %``A CoGeNT confirmation of the DAMA signal,''
  Phys.\ Lett.\ B {\bf 692} (2010) 65
  [arXiv:1004.1424 [hep-ph]].
  %%CITATION = ARXIV:1004.1424;%%

\bibitem{Chang:2010yk}
  S.~Chang, J.~Liu, A.~Pierce, N.~Weiner and I.~Yavin,
  %``CoGeNT Interpretations,''
  JCAP {\bf 1008} (2010) 018
  [arXiv:1004.0697 [hep-ph]].
  %%CITATION = JCAPA,1008,018;%%

\bibitem{Fitzpatrick:2010em}
  A.~L.~Fitzpatrick, D.~Hooper and K.~M.~Zurek,
  %``Implications of CoGeNT and DAMA for Light WIMP Dark Matter,''
  Phys.\ Rev.\  D {\bf 81} (2010) 115005
  [arXiv:1003.0014 [hep-ph]].
  %%CITATION = PHRVA,D81,115005;%%

\bibitem{Kopp:2009qt}
  J.~Kopp, T.~Schwetz and J.~Zupan,
  %``Global interpretation of direct Dark Matter searches after CDMS-II
  %results,''
  JCAP {\bf 1002} (2010) 014
  [updated version in arXiv:0912.4264 [hep-ph]].
  %%CITATION = JCAPA,1002,014;%%
  
\bibitem{Ahmed:2009zw}
  Z.~Ahmed {\it et al.} [CDMS-II Collaboration],
  %``Dark Matter Search Results from the CDMS II Experiment,''
  Science {\bf 327}, (2010),  1619-1621.
  [arXiv:{0912.3592} [astro-ph.CO]].

\bibitem{Aprile:2011hi}
  E.~Aprile {\it et al.}  [XENON100 Collaboration],
  %``Dark Matter Results from 100 Live Days of XENON100 Data,''
  Phys.\ Rev.\ Lett.\  {\bf 107} (2011) 131302
  [arXiv:1104.2549 [astro-ph.CO]].
  %%CITATION = ARXIV:1104.2549;%%

\bibitem{Archambault:2012pm}
  S.~Archambault {\it et al.}  [PICASSO Collaboration],
  %``Constraints on Low-Mass WIMP Interactions on $^{19}F$ from PICASSO,''
  arXiv:1202.1240 [hep-ex].
  %%CITATION = ARXIV:1202.1240;%%

\bibitem{Bernabei:2008jm}
  R.~Bernabei, P.~Belli, A.~Incicchitti and D.~Prosperi,
  %``Liquid Noble gases for Dark Matter searches: a synoptic survey,''
  arXiv:0806.0011 [astro-ph].
  %%CITATION = ARXIV:0806.0011;%%

\bibitem{Collar:2011wq}
  J.~I.~Collar,
  %``A Realistic Assessment of the Sensitivity of XENON10 and XENON100 to Light-Mass WIMPs,''
  arXiv:1106.0653 [astro-ph.CO].
  %%CITATION = ARXIV:1106.0653;%%

\bibitem{Collar:2011kf}
  J.~I.~Collar,
  %``A comparison between the low-energy spectra from CoGeNT and CDMS,''
  arXiv:1103.3481 [astro-ph.CO].
  %%CITATION = ARXIV:1103.3481;%%

\bibitem{Bernabei:2007hw}
  R.~Bernabei, P.~Belli, F.~Montecchia, F.~Nozzoli, F.~Cappella, A.~Incicchitti, D.~Prosperi and R.~Cerulli {\it et al.},
  %``Possible implications of the channeling effect in NaI(Tl) crystals,''
  Eur.\ Phys.\ J.\ C {\bf 53} (2008) 205
  [arXiv:0710.0288 [astro-ph]].
  %%CITATION = ARXIV:0710.0288;%%
  
\bibitem{Bozorgnia:2010xy}
  N.~Bozorgnia, G.~B.~Gelmini and P.~Gondolo,
  %``Channeling in direct dark matter detection I: channeling fraction in NaI (Tl) crystals,''
  JCAP {\bf 1011} (2010) 019
  [arXiv:1006.3110 [astro-ph.CO]].
  %%CITATION = ARXIV:1006.3110;%%

\bibitem{Feldstein:2009np}
  B.~Feldstein, A.~L.~Fitzpatrick, E.~Katz and B.~Tweedie,
  %``A Simple Explanation for DAMA with Moderate Channeling,''
  JCAP {\bf 1003} (2010) 029
  [arXiv:0910.0007 [hep-ph]].
  %%CITATION = ARXIV:0910.0007;%%

%\cite{Bagnasco:1993st}
\bibitem{Bagnasco:1993st} 
  J.~Bagnasco, M.~Dine and S.~D.~Thomas,
  %``Detecting technibaryon dark matter,''
  Phys.\ Lett.\ B {\bf 320}, 99 (1994)
  [hep-ph/9310290].
  %%CITATION = HEP-PH/9310290;%%

%\cite{Pospelov:2000bq}
\bibitem{Pospelov:2000bq}
  M.~Pospelov and T.~ter Veldhuis,
  %``Direct and indirect limits on the electromagnetic form-factors of WIMPs,''
ÊÊPhys.\ Lett.\ B {\bf 480} (2000) 181
ÊÊ[hep-ph/0003010].
ÊÊ%%CITATION = HEP-PH/0003010;%%

\bibitem{Sigurdson:2004zp}
  K.~Sigurdson, M.~Doran, A.~Kurylov, R.~R.~Caldwell and M.~Kamionkowski,
  %``Dark-matter electric and magnetic dipole moments,''
  Phys.\ Rev.\ D {\bf 70} (2004) 083501
   [Erratum-ibid.\ D {\bf 73} (2006) 089903]
  [astro-ph/0406355].
  %%CITATION = ASTRO-PH/0406355;%%

\bibitem{Barger:2010gv} 
  V.~Barger, W.~-Y.~Keung and D.~Marfatia,
  %``Electromagnetic properties of dark matter: Dipole moments and charge form factor,''
  Phys.\ Lett.\ B {\bf 696}, 74 (2011)
  [arXiv:1007.4345 [hep-ph]].
  %%CITATION = ARXIV:1007.4345;%%

\bibitem{Chang:2010en}
  S.~Chang, N.~Weiner and I.~Yavin,
  %``Magnetic Inelastic Dark Matter,''
  Phys.\ Rev.\ D {\bf 82} (2010) 125011
  [arXiv:1007.4200 [hep-ph]].
  %%CITATION = ARXIV:1007.4200;%%

\bibitem{Cho:2010br}
  W.~S.~Cho, J.~-H.~Huh, I.~-W.~Kim, J.~E.~Kim and B.~Kyae,
  %``Constraining WIMP magnetic moment from CDMS II experiment,''
  Phys.\ Lett.\ B {\bf 687} (2010) 6
   [Erratum-ibid.\ B {\bf 694} (2011) 496]
  [arXiv:1001.0579 [hep-ph]].
  %%CITATION = ARXIV:1001.0579;%%

\bibitem{Heo:2009vt}
  J.~H.~Heo,
  %``Minimal Dirac Fermionic Dark Matter with Nonzero Magnetic Dipole Moment,''
  Phys.\ Lett.\ B {\bf 693} (2010) 255
  [arXiv:0901.3815 [hep-ph]].
  %%CITATION = ARXIV:0901.3815;%%

\bibitem{Gardner:2008yn}
  S.~Gardner,
  %``Shedding Light on Dark Matter: A Faraday Rotation Experiment to Limit a Dark Magnetic Moment,''
  Phys.\ Rev.\ D {\bf 79} (2009) 055007
  [arXiv:0811.0967 [hep-ph]].
  %%CITATION = ARXIV:0811.0967;%%

\bibitem{Masso:2009mu}
  E.~Masso, S.~Mohanty and S.~Rao,
  %``Dipolar Dark Matter,''
  Phys.\ Rev.\ D {\bf 80} (2009) 036009
  [arXiv:0906.1979 [hep-ph]].
  %%CITATION = ARXIV:0906.1979;%%

\bibitem{Banks:2010eh}
  T.~Banks, J.~-F.~Fortin and S.~Thomas,
  %``Direct Detection of Dark Matter Electromagnetic Dipole Moments,''
  arXiv:1007.5515 [hep-ph].
  %%CITATION = ARXIV:1007.5515;%%

\bibitem{Fortin:2011hv}
  J.~-F.~Fortin and T.~M.~P.~Tait,
  %``Collider Constraints on Dipole-Interacting Dark Matter,''
  Phys.\ Rev.\ D {\bf 85} (2012) 063506
  [arXiv:1103.3289 [hep-ph]].
  %%CITATION = ARXIV:1103.3289;%%

\bibitem{Barger:2012pf}
  V.~Barger, W.~-Y.~Keung, D.~Marfatia and P.~-Y.~Tseng,
  %``Dipole Moment Dark Matter at the LHC,''
  arXiv:1206.0640 [hep-ph].
  %%CITATION = ARXIV:1206.0640;%%

\bibitem{Cline:2012is} 
  J.~M.~Cline, Z.~Liu and W.~Xue,
  %``Millicharged Atomic Dark Matter,''
ÊÊarXiv:1201.4858 [hep-ph].
ÊÊ%%CITATION = ARXIV:1201.4858;%%

\bibitem{Foadi:2007ue}
  R.~Foadi, M.~T.~Frandsen, T.~A.~Ryttov and F.~Sannino,
  %``Minimal Walking Technicolor: Set Up for Collider Physics,''
  Phys.\ Rev.\ D {\bf 76} (2007) 055005
  [arXiv:0706.1696 [hep-ph]].
  %%CITATION = ARXIV:0706.1696;%%

\bibitem{Ryttov:2008xe}
  T.~A.~Ryttov and F.~Sannino,
  %``Ultra Minimal Technicolor and its Dark Matter TIMP,''
  Phys.\ Rev.\ D {\bf 78} (2008) 115010
  [arXiv:0809.0713 [hep-ph]].
  %%CITATION = ARXIV:0809.0713;%%

\bibitem{Sannino:2008nv} 
  F.~Sannino and R.~Zwicky,
  %``Unparticle and Higgs as Composites,''
  Phys.\ Rev.\ D {\bf 79}, 015016 (2009)
  [arXiv:0810.2686 [hep-ph]].
  %%CITATION = ARXIV:0810.2686;%%

\bibitem{Sannino:2009za}
  F.~Sannino,
  %``Conformal Dynamics for TeV Physics and Cosmology,''
  Acta Phys.\ Polon.\ B {\bf 40} (2009) 3533
  [arXiv:0911.0931 [hep-ph]].
  %%CITATION = ARXIV:0911.0931;%%

\bibitem{Nardi:2008ix} 
  E.~Nardi, F.~Sannino and A.~Strumia,
  %``Decaying Dark Matter can explain the e+- excesses,''
  JCAP {\bf 0901}, 043 (2009)
  [arXiv:0811.4153 [hep-ph]].
  %%CITATION = ARXIV:0811.4153;%%

\bibitem{Lewis:2011zb} 
  R.~Lewis, C.~Pica and F.~Sannino,
  %``Light Asymmetric Dark Matter on the Lattice: SU(2) Technicolor with Two Fundamental Flavors,''
  Phys.\ Rev.\ D {\bf 85}, 014504 (2012)
  [arXiv:1109.3513 [hep-ph]].
  %%CITATION = ARXIV:1109.3513;%%

\bibitem{Fornengo:2011sz}
  N.~Fornengo, P.~Panci and M.~Regis,
  %``Long-Range Forces in Direct Dark Matter Searches,''
  Phys.\ Rev.\ D {\bf 84} (2011) 115002
  [arXiv:1108.4661 [hep-ph]].
  %%CITATION = ARXIV:1108.4661;%%

\bibitem{Chun:2010ve}
  E.~J.~Chun, J.~-C.~Park, S.~Scopel,
  %``Dark matter and a new gauge boson through kinetic mixing,''
  JHEP {\bf 1102 } (2011)  100.
  [arXiv:1011.3300 [hep-ph]].

\bibitem{DelNobile:2011uf}
  E.~Del Nobile and F.~Sannino,
  %``Dark Matter Effective Theory,''
  arXiv:1102.3116 [hep-ph]. Accepted for publication.
  %%CITATION = ARXIV:1102.3116;%% 

\bibitem{Foadi:2008qv}
  R.~Foadi, M.~T.~Frandsen and F.~Sannino,
  %``Technicolor Dark Matter,''
  Phys.\ Rev.\ D {\bf 80} (2009) 037702
  [arXiv:0812.3406 [hep-ph]].
  %%CITATION = ARXIV:0812.3406;%%

\bibitem{Frandsen:2009mi}
  M.~T.~Frandsen and F.~Sannino,
  %``iTIMP: isotriplet Technicolor Interacting Massive Particle as Dark Matter,''
  Phys.\ Rev.\ D {\bf 81} (2010) 097704
  [arXiv:0911.1570 [hep-ph]].
  %%CITATION = ARXIV:0911.1570;%%

\bibitem{PDG}
\href{http://pdg.lbl.gov/}{Particle Data Group} webpage of \href{http://pdg.lbl.gov/2011/AtomicNuclearProperties/index.html}{Atomic and Nuclear Properties}.

\bibitem{IAEA}
N.~J.~Stone, \href{http://www-nds.iaea.org/publications/indc/indc-nds-0594.pdf}{``Table of Nuclear Magnetic Dipole and Electric Quadrupole Moments''} released by the IAEA \href{http://www-nds.iaea.org/publications/indc/indc-nds-0594/}{Nuclear Data Services}.

\bibitem{Fitzpatrick:2012ix}
  A.~L.~Fitzpatrick, W.~Haxton, E.~Katz, N.~Lubbers and Y.~Xu,
  %``The Effective Field Theory of Dark Matter Direct Detection,''
  arXiv:1203.3542 [hep-ph].
  %%CITATION = ARXIV:1203.3542;%%

\bibitem{Helm}
R.~H.~Helm,
Phys. \ Rev. \ {\bf 104}, (1956), 1466.

\bibitem{Fornengo:2003fm}
  N.~Fornengo and S.~Scopel,
  %``Temporal distortion of annual modulation at low recoil energies,''
  Phys.\ Lett.\ B {\bf 576} (2003) 189
  [hep-ph/0301132].
  %%CITATION = HEP-PH/0301132;%%

\bibitem{DelNobile:2011yb}
  E.~Del Nobile, C.~Kouvaris, F.~Sannino and J.~Virkajarvi,
  %``Dark Matter Interference,''
  arXiv:1111.1902 [hep-ph].
  %%CITATION = ARXIV:1111.1902;%%

\bibitem{Gao:2011ka}
  X.~Gao, Z.~Kang, T.~Li,
  %``Light Dark Matter Models with Isospin Violation,''
  [arXiv:1107.3529 [hep-ph]].  

\bibitem{Pato:2011de}
  M.~Pato,
  %``What can(not) be measured with ton-scale dark matter direct detection experiments,''
  JCAP {\bf 1110} (2011) 035
  [arXiv:1106.0743 [astro-ph.CO]].
  %%CITATION = ARXIV:1106.0743;%%
  
\bibitem{Chen:2011vd}
  S.~-L.~Chen and Y.~Zhang,
  %``Isospin-Violating Dark Matter and Neutrinos From the Sun,''
  Phys.\ Rev.\ D {\bf 84} (2011) 031301
  [arXiv:1106.4044 [hep-ph]].
  %%CITATION = ARXIV:1106.4044;%%

\bibitem{Frandsen:2011cg}
  M.~T.~Frandsen, F.~Kahlhoefer, S.~Sarkar and K.~Schmidt-Hoberg,
  %``Direct detection of dark matter in models with a light Z',''
  JHEP {\bf 1109} (2011) 128
  [arXiv:1107.2118 [hep-ph]].
  %%CITATION = ARXIV:1107.2118;%%

\bibitem{Kurylov:2003ra}
  A.~Kurylov, M.~Kamionkowski,
  %``Generalized analysis of weakly interacting massive particle searches,''
  Phys.\ Rev.\  {\bf D69 } (2004)  063503.
  [hep-ph/0307185].
  
\bibitem{Cline:2011zr}
  J.~M.~Cline and A.~R.~Frey,
  %``Minimal hidden sector models for CoGeNT/DAMA events,''
  Phys.\ Rev.\ D {\bf 84} (2011) 075003
  [arXiv:1108.1391 [hep-ph]].
  %%CITATION = ARXIV:1108.1391;%%
  
\bibitem{Cline:2011uu}
  J.~M.~Cline and A.~R.~Frey,
  %``Light dark matter versus astrophysical constraints,''
  Phys.\ Lett.\ B {\bf 706} (2012) 384
  [arXiv:1109.4639 [hep-ph]].
  %%CITATION = ARXIV:1109.4639;%%

\bibitem{Bernabei:2008yh}
  R.~Bernabei {\it et al.}  [DAMA Collaboration],
  %``The DAMA/LIBRA apparatus,''
  Nucl.\ Instrum.\ Meth.\  {\bf A592}, (2008), 297,
  [arXiv:{0804.2738} [astro-ph]].
  %%CITATION = NUIMA,A592,297;%%

\bibitem{Bernabei:1996vj}
  R.~Bernabei, P.~Belli, V.~Landoni, F.~Montecchia, W.~Di Nicolantonio, A.~Incicchitti, D.~Prosperi and C.~Bacci {\it et al.},
  %``New limits on WIMP search with large-mass low-radioactivity NaI(Tl) set-up at Gran Sasso,''
  Phys.\ Lett.\ B {\bf 389} (1996) 757.
  %%CITATION = PHLTA,B389,757;%%

\bibitem{Aalseth:2008rx}
  C.~E.~Aalseth {\it et al.} [CoGeNT Collaboration],
  %``Experimental constraints on a dark matter origin for the DAMA annual modulation effect,''
  Phys.\ Rev.\ Lett.\  {\bf 101}, (2008),  251301.
  [arXiv:{0807.0879} [astro-ph]].

\bibitem{Barbeau:2007qi}
  P.~S.~Barbeau, J.~I.~Collar and O.~Tench,
  %``Large-Mass Ultra-Low Noise Germanium Detectors: Performance and
  %Applications in Neutrino and Astroparticle Physics,''
  JCAP {\bf 0709}, (2007), 009,
  [arXiv:{nucl-ex/0701012}].
  %%CITATION = JCAPA,0709,009;%%
  
\bibitem{Ahmed:2010hw}
  Z.~Ahmed {\it et al.}  [CDMS-II and CDMS Collaborations],
  %``Search for inelastic dark matter with the CDMS II experiment,''
  Phys.\ Rev.\ D {\bf 83} (2011) 112002
  [arXiv:1012.5078 [astro-ph.CO]].
  %%CITATION = ARXIV:1012.5078;%%
  
\bibitem{Ahmed:2011gh}
  Z.~Ahmed {\it et al.}  [CDMS and EDELWEISS Collaborations],
  %``Combined Limits on WIMPs from the CDMS and EDELWEISS Experiments,''
  Phys.\ Rev.\ D {\bf 84} (2011) 011102
  [arXiv:1105.3377 [astro-ph.CO]].
  %%CITATION = ARXIV:1105.3377;%%

\bibitem{Aprile:2011hx}
  E.~Aprile {\it et al.}  [XENON100 Collaboration],
  %``Likelihood Approach to the First Dark Matter Results from XENON100,''
  Phys.\ Rev.\ D {\bf 84} (2011) 052003
  [arXiv:1103.0303 [hep-ex]].
  %%CITATION = ARXIV:1103.0303;%%

\bibitem{PICASSO1}
R. E. Apfel, Nucl. Inst. and Meth. 162 (1979) 603--608.

\bibitem{PICASSO2}
H. Ing, R. Noulty, T. McLean, Radiation Measurements 27 (1997) 1--11.

\bibitem{Komatsu:2010fb}
  E.~Komatsu {\it et al.}  [WMAP Collaboration],
  %``Seven-Year Wilkinson Microwave Anisotropy Probe (WMAP) Observations: Cosmological Interpretation,''
  Astrophys.\ J.\ Suppl.\  {\bf 192} (2011) 18
  [arXiv:1001.4538 [astro-ph.CO]].
  %%CITATION = ARXIV:1001.4538;%%

\bibitem{Gondolo:1990dk}
  P.~Gondolo and G.~Gelmini,
  %``Cosmic abundances of stable particles: Improved analysis,''
  Nucl.\ Phys.\ B {\bf 360} (1991) 145.
  %%CITATION = NUPHA,B360,145;%%

\bibitem{Belyaev:2010kp} 
  A.~Belyaev, M.~T.~Frandsen, S.~Sarkar and F.~Sannino,
  %``Mixed dark matter from technicolor,''
  Phys.\ Rev.\ D {\bf 83}, 015007 (2011)
  [arXiv:1007.4839 [hep-ph]].
  %%CITATION = ARXIV:1007.4839;%%

%\cite{Cirelli:2011ac}
\bibitem{Cirelli:2011ac} 
  M.~Cirelli, P.~Panci, G.~Servant and G.~Zaharijas,
  %``Consequences of DM/antiDM Oscillations for Asymmetric WIMP Dark Matter,''
  JCAP {\bf 1203}, 015 (2012)
  [arXiv:1110.3809 [hep-ph]].
  %%CITATION = ARXIV:1110.3809;%%

%\cite{Tulin:2012re}
\bibitem{Tulin:2012re} 
  S.~Tulin, H.~-B.~Yu and K.~M.~Zurek,
  %``Oscillating Asymmetric Dark Matter,''
  arXiv:1202.0283 [hep-ph].
  %%CITATION = ARXIV:1202.0283;%%

\bibitem{Buckley:2011ye} 
  M.~R.~Buckley and S.~Profumo,
  %``Regenerating a Symmetry in Asymmetric Dark Matter,''
  Phys.\ Rev.\ Lett.\  {\bf 108}, 011301 (2012)
  [arXiv:1109.2164 [hep-ph]].
  %%CITATION = ARXIV:1109.2164;%%

\bibitem{Galli:2009zc}
  S.~Galli, F.~Iocco, G.~Bertone and A.~Melchiorri,
  %``CMB constraints on Dark Matter models with large annihilation cross-section,''
  Phys.\ Rev.\ D {\bf 80} (2009) 023505
  [arXiv:0905.0003 [astro-ph.CO]].
  %%CITATION = ARXIV:0905.0003;%%

\bibitem{Slatyer:2009yq}
  T.~R.~Slatyer, N.~Padmanabhan and D.~P.~Finkbeiner,
  %``CMB Constraints on WIMP Annihilation: Energy Absorption During the Recombination Epoch,''
  Phys.\ Rev.\ D {\bf 80} (2009) 043526
  [arXiv:0906.1197 [astro-ph.CO]].
  %%CITATION = ARXIV:0906.1197;%%

\bibitem{Huetsi:2009ex}
  G.~Huetsi, A.~Hektor and M.~Raidal,
  %``Constraints on leptonically annihilating Dark Matter from reionization and extragalactic gamma background,''
  Astron.\ Astrophys.\  {\bf 505} (2009) 999
  [arXiv:0906.4550 [astro-ph.CO]].
  %%CITATION = ARXIV:0906.4550;%%

\bibitem{Cirelli:2009bb}
  M.~Cirelli, F.~Iocco and P.~Panci,
  %``Constraints on Dark Matter annihilations from reionization and heating of the intergalactic gas,''
  JCAP {\bf 0910} (2009) 009
  [arXiv:0907.0719 [astro-ph.CO]].
  %%CITATION = ARXIV:0907.0719;%%

\bibitem{Hutsi:2011vx}
  G.~Hutsi, J.~Chluba, A.~Hektor and M.~Raidal,
  %``WMAP7 and future CMB constraints on annihilating dark matter: implications on GeV-scale WIMPs,''
  Astron.\ Astrophys.\  {\bf 535} (2011) A26
  [arXiv:1103.2766 [astro-ph.CO]].
  %%CITATION = ARXIV:1103.2766;%%

\bibitem{Galli:2011rz}
  S.~Galli, F.~Iocco, G.~Bertone and A.~Melchiorri,
  %``Updated CMB constraints on Dark Matter annihilation cross-sections,''
  Phys.\ Rev.\ D {\bf 84} (2011) 027302
  [arXiv:1106.1528 [astro-ph.CO]].
  %%CITATION = ARXIV:1106.1528;%%

\bibitem{Ackermann:2011wa}
  M.~Ackermann {\it et al.}  [Fermi-LAT Collaboration],
  %``Constraining Dark Matter Models from a Combined Analysis of Milky Way Satellites with the Fermi Large Area Telescope,''
  Phys.\ Rev.\ Lett.\  {\bf 107} (2011) 241302
  [arXiv:1108.3546 [astro-ph.HE]].
  %%CITATION = ARXIV:1108.3546;%%
  %See also: talk by Maja L. Garde at the \myurl{fermi.gsfc.nasa.gov/science/symposium/2011/program/}{FERMI Symposium 2011}.

\bibitem{Cirelli:2009dv}
  M.~Cirelli, P.~Panci and P.~D.~Serpico,
  %``Diffuse gamma ray constraints on annihilating or decaying Dark Matter after Fermi,''
  Nucl.\ Phys.\ B {\bf 840} (2010) 284
  [arXiv:0912.0663 [astro-ph.CO]].
  %%CITATION = ARXIV:0912.0663;%%

\bibitem{Papucci:2009gd}
  M.~Papucci and A.~Strumia,
  %``Robust implications on Dark Matter from the first FERMI sky gamma map,''
  JCAP {\bf 1003} (2010) 014
  [arXiv:0912.0742 [hep-ph]].
  %%CITATION = ARXIV:0912.0742;%%

\bibitem{Zaharijas:2010ca}
  G.~Zaharijas {\it et al.}  [for the Fermi-LAT Collaboration],
  %``Constraints on the Galactic Halo Dark Matter from Fermi-LAT Diffuse Measurements,''
  PoS IDM {\bf 2010} (2011) 111
  [arXiv:1012.0588 [astro-ph.HE]].
  %%CITATION = ARXIV:1012.0588;%%

\bibitem{Abdo:2010nc}
  A.~A.~Abdo %, M.~Ackermann, M.~Ajello, W.~B.~Atwood, L.~Baldini, J.~Ballet, G.~Barbiellini and D.~Bastieri 
  {\it et al.},
  %``Fermi LAT Search for Photon Lines from 30 to 200 GeV and Dark Matter Implications,''
  Phys.\ Rev.\ Lett.\  {\bf 104} (2010) 091302
  [arXiv:1001.4836 [astro-ph.HE]].
  %%CITATION = ARXIV:1001.4836;%%

\bibitem{Morselli}
Talk given by A.~Morselli at the \href{http://kitpc.itp.ac.cn/scw.jsp?id=WD20110926&i=sched}{7th International workshop on Dark Side of the Universe}%\href{http://kitpc.itp.ac.cn/files/activities/WD20110926/report/DSU2011-A.Morselli.pdf}{talk}
.

\bibitem{Goodman:2010qn}
  J.~Goodman, M.~Ibe, A.~Rajaraman, W.~Shepherd, T.~M.~P.~Tait and H.~-B.~Yu,
  %``Gamma Ray Line Constraints on Effective Theories of Dark Matter,''
  Nucl.\ Phys.\ B {\bf 844} (2011) 55
  [arXiv:1009.0008 [hep-ph]].
  %%CITATION = ARXIV:1009.0008;%%

\bibitem{Kouvaris:2011fi} 
  C.~Kouvaris and P.~Tinyakov,
  %``Excluding Light Asymmetric Bosonic Dark Matter,''
  Phys.\ Rev.\ Lett.\  {\bf 107}, 091301 (2011)
  [arXiv:1104.0382 [astro-ph.CO]].
  %%CITATION = ARXIV:1104.0382;%%

\bibitem{McDermott:2011jp} 
  S.~D.~McDermott, H.~-B.~Yu and K.~M.~Zurek,
  %``Constraints on Scalar Asymmetric Dark Matter from Black Hole Formation in Neutron Stars,''
  Phys.\ Rev.\ D {\bf 85}, 023519 (2012)
  [arXiv:1103.5472 [hep-ph]].
  %%CITATION = ARXIV:1103.5472;%%  

\bibitem{Guver:2012ba} 
  T.~Guver, A.~E.~Erkoca, M.~H.~Reno and I.~Sarcevic,
  %``On the capture of dark matter by neutron stars,''
  arXiv:1201.2400 [hep-ph].
  %%CITATION = ARXIV:1201.2400;%%  

\bibitem{Kouvaris:2011gb} 
  C.~Kouvaris,
  %``Limits on Self-Interacting Dark Matter,''
  arXiv:1111.4364 [astro-ph.CO].
  %%CITATION = ARXIV:1111.4364;%%

\bibitem{Kouvaris:2010jy} 
  C.~Kouvaris and P.~Tinyakov,
  %``Constraining Asymmetric Dark Matter through observations of compact stars,''
  Phys.\ Rev.\ D {\bf 83}, 083512 (2011)
  [arXiv:1012.2039 [astro-ph.HE]].
  %%CITATION = ARXIV:1012.2039;%%

\end{thebibliography}
\end{document}